\pdfoutput=1

\documentclass[11pt,a4paper]{article}
\usepackage{array}
\usepackage{bm}
\usepackage{booktabs}
\usepackage{jheppub}
\usepackage{bbm}
\usepackage{mathrsfs}
\usepackage{mathtools}

\let\de=\partial
\let\a=\alpha
\let\b=\beta
\let\c=\chi
\let\g=\gamma
\let\d=\delta
\let\eps=\epsilon
\let\k=\kappa
\let\l=\lambda
\let\m=\mu
\let\n=\nu
\let\p=\phi
\let\om=\omega
\let\Om=\Omega
\let\S=\Sigma
\let\w=\wedge

\newcommand{\im}{\text{i}}
\newcommand{\gr}[1]{\mathrm{#1}}
\newcommand{\La}{\mathscr{L}}
\newcommand{\dd}{\text{d}}
\DeclareMathOperator{\ho}{\star}
\DeclareMathOperator{\tr}{tr}
\DeclareMathOperator{\sgn}{sgn}
\newcommand{\vek}[1]{\bm{#1}}
\newcommand{\A}{\mathfrak a}
\newcommand{\B}{\mathfrak b}
\newcommand{\I}{\mathcal{I}}
\newcommand{\J}{\mathcal{J}}
\newcommand{\K}{\mathcal{K}_\text{c.t.}}
\newcommand{\CP}{\mathbb{CP}}
\newcommand{\R}{\mathbb{R}}
\renewcommand{\deg}[2]{#1^{(#2)}}
\newcommand{\codeg}[2]{#1^{[#2]}}

\title{Field theories with higher-group symmetry\\ from composite currents}

\author{Tom\'a\v{s} Brauner}
\affiliation{Department of Mathematics and Physics, University of Stavanger,\\
N-4036 Stavanger, Norway}
\emailAdd{tomas.brauner@uis.no}

\abstract{Higher-form symmetries are associated with transformations that only act on extended objects, not on point particles. Typically, higher-form symmetries live alongside ordinary, point-particle (0-form), symmetries and they can be jointly described in terms of a direct product symmetry group. However, when the actions of 0-form and higher-form symmetries become entangled, a more general mathematical structure is required, related to higher categorical groups. Systems with continuous higher-group symmetry were previously constructed in a top-down manner, descending from quantum field theories with a specific mixed 't Hooft anomaly. I show that higher-group symmetry also naturally emerges from a bottom-up, low-energy perspective, when the physical system at hand contains at least two different given, spontaneously broken symmetries. This leads generically to a hierarchy of emergent higher-form symmetries, corresponding to the Grassmann algebra of topological currents of the theory, with an underlying higher-group structure. Examples of physical systems featuring such higher-group symmetry include superfluid mixtures and variants of axion electrodynamics.}


\arxivnumber{2012.00051}

\begin{document}
 
\maketitle


\section{Introduction}
\label{sec:intro}

The notion of symmetry is central to our understanding of the laws of nature. Since more than a century ago, symmetries have been linked to conservation laws by the celebrated first Noether theorem; see ref.~\cite{Kosmann-Schwarzbach2011a} for a historical account of the subject. Numerous generalizations of the original approach due to Noether, based on local symmetries\footnote{In mathematical literature, it is common to use the term ``local symmetry'' for a set of transformations depending only on the values of the spacetime coordinates, fields and a finite number of their derivatives at the given spacetime point. This is not to be confused with the notion of local symmetry in classical and quantum field theory where the parameters of the transformation are, in contrast to global symmetries, functions of the spacetime coordinates.} of a variational problem, have emerged throughout the years and proven of importance to both mathematics and physics. Thus, for instance, the concepts of a local symmetry and a local conservation law have been lifted to the more general partial differential equation setting, and their relationship therein has been established~\cite{Olver1986a,Bluman2002a}.

About thirty years ago, the study of generalized local conservation laws, based on higher-rank antisymmetric tensor currents, was initiated in parallel in the mathematics~\cite{Zharinov1992a,Bryant1995a} and physics~\cite{Barnich1995a,Anderson1996a,Barnich2000a} communities. These have become known as ``lower-degree conservation laws,'' owing to their description in terms of closed differential forms of degree lower than $D-1$ in a spacetime of dimension $D$; closed $(D-1)$-forms are Hodge-dual to the vector currents of ordinary conservation laws. The concept of a lower-degree conservation law is, in fact, closely related to that of ``nonlocal symmetry'' of a partial differential equation, whose history goes further back, at least to mid-1980s~\cite{Krasilshchik1984a,Vladimirov1990a,Akhatov1991a}, see also ref.~\cite{Bluman2010a} for a pedagogical introduction aimed at practical applications.

In the last decade or so, lower-degree symmetries have been attracting growing interest due to their importance for understanding the ordered phases of quantum matter, see refs.~\cite{Nussinov2009a,Gaiotto2015a,Sharpe2015a} for some early works. In the high-energy physics community, it is now common to use the term ``higher-form'' rather than ``lower-degree'' symmetry, which will be used consistently throughout the rest of this paper. In particular the work of Gaiotto et al.~\cite{Gaiotto2015a} anticipated much of the subsequent developments, including the perimeter law for Wilson surfaces as a criterion for spontaneous breaking of higher-form symmetries~\cite{Hofman2019a}, higher-form generalizations of the Goldstone~\cite{Hofman2019a,Lake2018a} and Coleman-Mermin-Wagner~\cite{Lake2018a} theorems, and the importance of mixed 't Hooft anomalies for Landau's classification of phase transitions~\cite{Delacretaz2020a}. Thanks to the fact that local conservation laws govern small departures from thermodynamic equilibrium, continuous higher-form symmetries have found natural applications to magnetohydrodynamics~\cite{Grozdanov2017a,Armas2019a,Glorioso2018a,Armas2020a} and to the effective field theory (EFT) description of systems with topological defects~\cite{Grozdanov2018a,Armas2020b}. Finally, it has been pointed out~\cite{Sogabe2019a,Hidaka2020b} that the spectrum of Nambu-Goldstone (NG) bosons arising from spontaneously broken higher-form symmetries in nonrelativistic systems features subtleties, analogous to those found for spontaneously broken 0-form symmetries, see refs.~\cite{Watanabe2020a,Beekman2019a,AlvarezGaume2020a} for recent reviews.

Remarkably, this still does not exhaust the spectrum of possible realizations of symmetry. The actions of symmetries of different degrees may become entangled, in which case they can no longer by captured by a mere group. A more general mathematical structure is required, dubbed a ``higher group,'' see ref.~\cite{Baez2011a} for a nontechnical introduction. Discrete higher-group symmetries were observed for instance in refs.~\cite{Pantev2005a,Pantev2006a,Kapustin2013a,Tachikawa2020a}. Continuous 2-group symmetries were investigated thoroughly by C\'ordova et al.~\cite{Cordova2019a,Benini2019a}. To date, two distinct mechanisms how a continuous higher-group symmetry may arise have been identified. In ref.~\cite{Cordova2019a} it was shown how a global 2-group symmetry descends from a mixed 't Hooft anomaly of ordinary 0-form symmetries once part of the symmetry has been gauged. A higher-group symmetry may also arise from a topological coupling of otherwise disconnected sectors possessing symmetries of different degrees; this is the case for the 3-group symmetry of axion electrodynamics~\cite{Hidaka2020d,Hidaka2020a}, see also the very recent ref.~\cite{Brennan2020a}.

The goal of this paper is to show that there is another, apparently unrelated, mechanism whereby a continuous higher-group symmetry may emerge. I take the low-energy perspective, assuming that to start with, a given many-body system features spontaneously broken continuous $p$-form symmetry, and its low-energy physics is governed by a bosonic EFT for the ensuing NG mode(s). It has been noticed previously that for $p$-form $\gr{U(1)}$ symmetries, this automatically implies the existence of an emergent dual $(D-p-2)$-form symmetry that has a mixed 't Hooft anomaly with the original $p$-form symmetry~\cite{Gaiotto2015a,Delacretaz2020a}, see also ref.~\cite{Ji2020a} for a similar observation in case of a finite Abelian symmetry group. The new element added here is that when at least two such pairs of symmetries are present, one can construct new composite, topologically conserved currents. In principle, the topological currents of the system then span a whole hierarchy of symmetries of different degrees, which turns out to naturally support a higher-group symmetry structure. This higher-group structure can be probed by coupling the system to a set of background gauge fields, one for each conserved current.


\subsection{Some notation and terminology, and the main idea}
\label{subsec:mainidea}

Throughout the text, I use systematically the language of differential forms. This allows one among others to put under the same umbrella of ``$p$-form superfluids'' systems as different as ordinary, $s$-wave superfluids and Maxwell's electrodynamics. It also underlines the geometric nature of all the symmetries involved, and considerably simplifies the notation without having to choose a specific spacetime dimension, $D$. Readers not familiar with differential forms may find an elementary introduction at a junior graduate level in ref.~\cite{Stone2009a}. A careful discussion of $p$-form superfluids without differential forms, including their higher-form symmetries, was given recently in ref.~\cite{Yamamoto2020a}.

For the sake of transparency, at the cost of cluttering somewhat the notation, I will indicate the degrees of all differential forms using superscripts. Thus, a $p$-form $\om$ will be denoted by $\deg\om p$; the only exception will be made for 0-form variables, for which the explicit superscript $\deg{}0$ will be dropped. It will frequently be more economic to indicate the codegree of the form. Hence, $\codeg\om p$ will denote a form of codegree $p$, or equivalently of degree $D-p$. The Hodge star operator will be defined in a chosen coordinate basis by
\begin{equation}
\ho(\dd x^{\mu_1}\w\dotsb\w\dd x^{\mu_p})\equiv\frac{\sqrt g}{(D-p)!}g^{\mu_1\nu_1}\dotsb g^{\mu_p\nu_p}\eps_{\nu_1\dotsb\nu_D}\dd x^{\nu_{p+1}}\w\dotsb\w\dd x^{\nu_D},
\end{equation}
where $g_{\m\n}$ is the spacetime metric and $g$ without subscripts its determinant, and $\eps_{\mu_1\dotsb\mu_D}$ is the fully antisymmetric Levi-Civita symbol defined so that $\eps_{1\dotsb D}=1$. On the few occasions it will be needed, I will implicitly assume that the spacetime metric has Euclidean signature so that the square of the Hodge star operator is fixed by
\begin{equation}
\ho\ho\deg\om p=(-1)^{p(D-p)}\deg\om p.
\end{equation}
A continuous $p$-form symmetry will be represented by a $(p+1)$-form current $\deg J{p+1}$. In the language of differential forms, current conservation takes the neat form
\begin{equation}
\dd^\dag\deg J{p+1}=0.
\label{conslaw}
\end{equation}
Here $\dd^\dag$ is the codifferential, defined by
\begin{equation}
\dd^\dag\deg\om p\equiv(-1)^{Dp+D+1}\ho\dd\ho\deg\om p.
\end{equation}
It is common to drop the overall sign and express the conservation law~\eqref{conslaw} as the closedness of the Hodge-dual current, $\ho\deg J{p+1}$.

For the purposes of this paper, I will distinguish two different types of conserved currents, which I will call ``Noether currents'' and ``topological currents.'' The conservation of Noether currents requires the equations of motion for the dynamical degrees of freedom to hold. Topological currents, on the other hand, are conserved identically. It is only the Noether currents that are in a correspondence with (local) symmetries of the action functional. Topological currents do not generate any transformation on the local field variables~\cite{Vyas2019a}, and are in fact considered trivial in the literature on symmetries of partial differential equations~\cite{Olver1986a}. There may, however, be finite-energy extended field configurations that carry an integral charge derived from such currents. This is typically the case when a symmetry of the system is spontaneously broken and the corresponding vacuum manifold has nontrivial topology, hence the term ``topological currents.''

The distinction between Noether and topological currents is not sharp. Indeed, it is usually possible to make a given Noether current trivially conserved by introducing a suitable dual field variable. This idea lies behind the concept of ``potential symmetries'' of partial differential equations~\cite{Bluman1988a,Cheviakov2010a,Cheviakov2010b}. In the low-energy EFT perspective employed in this paper, I will however assume that certain symmetry of the quantum many-body system at hand is \emph{given}, and is (partially or completely) spontaneously broken in the ground state. This given symmetry is responsible for the Noether currents of the EFT. There are standard techniques for the construction of effective actions once the symmetry-breaking pattern is known, at least for compact internal 0-form symmetries~\cite{Coleman1969a,Callan1969a,Watanabe2014a,Andersen2014a}. I will moreover assume that there are no other Noether currents in the effective theory; the existence of such currents would indicate that the symmetry-breaking pattern has not been identified properly. On top of the given Noether currents, the EFT may possess additional, accidental conserved currents; these are by construction topological.

To summarize, the separation between Noether and topological currents is fixed once the given symmetry is known \emph{and} the local degrees of freedom (NG bosons) are chosen. The conservation laws for Noether currents then naturally include the equations of motion for the NG bosons. Whatever topological currents are present, are emergent and descend directly from the geometry of the coset space on which the EFT lives.

I now finally get to the main idea of the paper. Suppose that our system possesses two conserved currents, $\deg{J}{p+1}$ and $\deg{K}{q+1}$, corresponding respectively to a $p$-form and a $q$-form symmetry. Then the exterior product of their Hodge duals, $(\ho\deg{J}{p+1})\w(\ho\deg{K}{q+1})$, is necessarily closed. Provided that $p+q>D-2$, it represents the Hodge dual of a new conserved current, corresponding to a new $(p+q-D+1)$-form symmetry. I will refer to $\deg{J}{p+1}$ and $\deg{K}{q+1}$ as ``primitive currents'' and to $\ho[(\ho\deg{J}{p+1})\w(\ho\deg{K}{q+1})]$ as a ``second-order composite current.'' This observation is trivially extended to the statement that all Hodge-dual currents of the system span a subalgebra of the Grassmann algebra of closed differential forms on the spacetime manifold. The corresponding set of conserved currents splits into primitive, second- and possibly higher-order composite currents.

The possibility of existence of composite currents places a constraint on the degrees of primitive currents one starts with. In concrete applications, Noether currents are usually 1-form or 2-form, making it impossible to construct new composite currents out of Noether currents alone in $D\geq4$ dimensions. On the other hand, recalling that a spontaneously broken $p$-form $\gr{U(1)}$ symmetry is associated with an emergent dual $(D-p-2)$-form symmetry, we see that it is natural to construct composite currents out of primitive \emph{topological} currents, which are as a rule of high degree. The typical picture one should expect is therefore an EFT with a given spontaneously broken symmetry, on top of which there exists a hierarchy of emergent symmetries, built upon the Grassmann algebra of closed Hodge-dual topological currents.\footnote{Such a hierarchy of emergent higher-form symmetries, describing conservation of topological defects in crystalline solids, was considered recently by Nissinen~\cite{Nissinen2020a}. An idea similar in spirit lies also behind the discussion of 2-group symmetries in six-dimensional quantum field theories in ref.~\cite{Cordova2020a}.}


\subsection{Outline of the paper}
\label{subsec:outline}

The rest of the paper starts with a brief introduction to higher-form symmetries in section~\ref{sec:review}. While a comprehensive but accessible review of the subject seems to be missing in the literature, it is definitely not the intention to provide such a review here. Instead, I will focus largely on the classical aspects of continuous higher-form symmetries that will be needed later. It is hopefully clear from the discussion above that the existence of topologically conserved currents is an essential ingredient of the mechanism behind higher-group symmetries, proposed in this paper. Section~\ref{sec:topcurrents} elaborates on when such currents can be expected, and how they depend on the geometry of the coset space of the given (Noether) symmetry. It turns out that for compact internal symmetries, there is a class of topological currents described by de Rham cohomology of the coset space, that is the same geometric structure that underlies the topological Wess-Zumino terms~\cite{DHoker1994a,DHoker1995b}.

The core of the paper comprises the following four sections. In section~\ref{sec:abelian} I first work out the details in the simplest possible setting, where all the given Noether symmetries are assumed to be Abelian, and only second-order composite currents are taken into account. Using the coupling of conserved currents to background gauge fields and the associated background gauge invariance, I demonstrate that a nontrivial higher-group structure is a necessary consequence of the existence of composite currents in the theory. Section~\ref{sec:abelianhigher} shows how the formalism generalizes to the whole hierarchy of composite currents; for the sake of notational simplicity it is assumed here that all the given symmetries are 0-form. Further insight into the nature of the new, composite symmetries is gained by switching to the dual picture where the roles of the given Noether symmetries and their dual topological symmetries are interchanged; this is discussed in section~\ref{sec:duality}. Finally, section~\ref{sec:nonabelian} presents the rather straightforward generalization that allows the given symmetries to be non-Abelian. Some further comments are deferred to the concluding section~\ref{sec:conclusions}.


\section{Short review of higher-form symmetries}
\label{sec:review}

It is customary to take the symmetry as a starting point and thence deduce the existence of a conservation law via Noether's theorem. For higher-form symmetries, the question how to formulate the action of the symmetry on a given set of local field variables is however subtle, and will only be touched superficially in section~\ref{sec:duality}. For the time being, I will take the pragmatic approach and start by postulating the existence of a conservation law for a $(p+1)$-form current,
\begin{equation}
\dd^\dagger\deg J{p+1}=0,\qquad
\text{or equivalently}\qquad
\de_\n(\sqrt g J^{\n\m_1\dotsb\m_p})=0,
\label{conservation}
\end{equation}
where $J^{\m_1\dotsb\m_{p+1}}$ is the fully antisymmetric contravariant tensor whose components are obtained from those of the $(p+1)$-form current $\deg J{p+1}$ by raising with the metric.

Writing eq.~\eqref{conservation} as $\dd\ho\deg J{p+1}=0$, one immediately arrives at the trivial corollary that for any closed form $\deg\om q$ where $0\leq q\leq p$, one has
\begin{equation}
\dd(\ho\deg J{p+1}\w\deg\om q)=0.
\end{equation}
Hence, any $p$-form symmetry implies via the associated conservation law~\eqref{conservation} the existence of an infinite number of lower-form local conservation laws. A simple example is provided by Maxwell's theory without matter in flat Minkowski spacetime, where the sourceless Maxwell equation $\dd^\dagger\deg F2=0$ gives a 1-form conservation law.\footnote{A larger set of nonlocal symmetries and conservation laws of Maxwell's equations in three and four spacetime dimensions was identified in refs.~\cite{Anco1997a,Anco2005a}.} Leaving for the moment the differential form notation, we find as an immediate consequence that for any scalar function $\p$, $\de_\m(F^{\m\n}\de_\n\p)=0$. The existence of such induced conservation laws is rarely mentioned, but has been noted explicitly for instance in refs.~\cite{Cheviakov2014a,Hofman2019a}.


\subsection{Integral charges}
\label{subsec:charges}

Associated with the local conservation law~\eqref{conservation} there is an integral charge, obtained by integrating the Hodge-dual current $\ho\deg J{p+1}$ over a surface of codimension $p+1$,
\begin{equation}
Q(\S_{D-p-1})\equiv\int_{\S_{D-p-1}}\ho\deg J{p+1}.
\label{charge}
\end{equation}
The surface $\S_{D-p-1}$ should typically be compact and without boundary, but may also be noncompact provided the asymptotic behavior of the current at infinity is constrained by a suitable boundary condition. Indeed, the general definition of the integral charge~\eqref{charge} reduces to its textbook version valid for ordinary, 0-form symmetries if one takes $\S_{D-1}$ as the spatial manifold --- whether the latter is compact or not --- embedded in the spacetime as a fixed time slice. The textbook conservation of integral charge then amounts to the invariance of $Q(\S_{D-1})$ under shifts of the spatial manifold in time.

Using the more geometric language of differential forms highlights the fact that charge conservation is a much more robust, topological concept. Namely, the charge $Q(\S_{D-p-1})$ is invariant under any smooth deformation of the surface $\S_{D-p-1}$, as long as the conservation law~\eqref{conservation} holds in the whole codimension-$p$ domain swept by the surface during the deformation. The charge can only change if the surface crosses a source of the current, which can be thought of as a $p$-dimensional defect on which the $p$-form $\dd^\dagger\deg{J}{p+1}$ is nonzero. In the same spirit, the integral charge $Q(\S_{D-p-1})$ necessarily vanishes in case $\S_{D-p-1}$ is closed and forms the boundary of some codimension-$p$ domain. Nonzero charge can only arise if $\S_{D-p-1}$ belongs to a nontrivial homology class, or is noncompact.

For a simple illustration of how the integral charge depends on the choice of surface to integrate over, refer again to Maxwell's theory without matter, for simplicity taken in the flat $D=4$ Minkowski spacetime. Here $\deg F2$ is the conserved current of a 1-form symmetry, so the integral charge is obtained in accord with eq.~\eqref{charge} by integrating $\ho\deg F2$ over some two-dimensional surface $\S_2$. If we choose $\S_2$ as a closed spatial surface, then $Q(\S_2)$ measures the electric flux through $\S_2$. For this reason, the 1-form symmetry associated with the conservation of $\deg F2$ is referred to as ``electric.'' The electric flux is indeed invariant under smooth deformations of the surface $\S_2$ thanks to the usual Gauss law of electrostatics. In line with the general discussion above, $Q(\S_2)$ can in fact only be nonzero if $\S_2$ belongs to a nontrivial homology class, which is effectively the case when some electric charge is present in the domain bounded by the surface. If, on the other hand, we choose $\S_2$ as the worldsheet of some fixed closed spatial curve $\g_1$, then $Q(\S_2)$ measures the circulation of the magnetic field around $\g_1$. In a static electromagnetic field, this is likewise invariant under smooth deformations of the curve. It can only be nonzero if $\g_1$ belongs to a nontrivial homology class, which is effectively the case when the curve is threaded by electric current, in accord with Amp\`ere's law.

Before concluding the discussion of integral charges, I will briefly mention an interesting nonrelativistic generalization of 1-form symmetries, put forward recently by Seiberg~\cite{Seiberg2020a}. In this generalization, the 1-form divergence $\dd^\dagger\deg J2$ vanishes only when acting on spatial vectors. This still makes it possible to define a conserved integral charge by integration over (spatial) codimension-1 spatial manifolds. Such charges are, however, no longer topological, that is, they change upon a deformation of the manifold. 


\subsection{Charged objects}
\label{subsec:objects}

The topological property of the integral charge $Q(\S_{D-p-1})$ can be promoted to an alternative definition of a global symmetry. In case an integral representation of the charge such as eq.~\eqref{charge} exists, this alternative definition is clearly equivalent to the current conservation property~\eqref{conservation}. Its advantage is, however, that it is more general. Namely, this line of reasoning leads to the construction of topological operators, realizing the symmetry on the Hilbert space of a given quantum system. The existence of such operators can be used to define even \emph{discrete} higher-form symmetries, where no local conserved current exists. For a nice review of the implementation of higher-form symmetries via quantum topological operators, the reader may consult ref.~\cite{Harlow2018a}.

Going back to continuous higher-form symmetries, the charge operator~\eqref{charge} generates the symmetry transformations on the given quantum system. For ordinary, 0-form symmetries, the transformations act naturally on local fields, and by extension on any nonlocal operator constructed out of local fields. The hallmark of a $p$-form symmetry is that the corresponding ``elementary'' charged objects\footnote{The terminology does not seem to be quite settled, and instead of ``charged object'' the term ``defect'' or simply ``operator'' is sometimes used.} are operators with support on some $p$-dimensional manifold $C_p$, which parallels the above observation that the classical source of the current $\deg J{p+1}$ can be thought of as a $p$-dimensional defect. The action of the charge operator~\eqref{charge} on the object is nontrivial if the surfaces $\S_{D-p-1}$ and $C_p$ are topologically linked in the spacetime manifold.

For a simple example, recall that for the electric 1-form symmetry of Maxwell's theory without matter in $D=4$ dimensions, the charge~\eqref{charge} measures electric flux when defined on a spatial surface $\S_2$. The corresponding charged object can be taken as the worldline of an electrically charged point particle that lies inside the domain bounded by $\S_2$. In the language of gauge theory, the charged operator is the Wilson line defined on the worldline of the particle.

Maxwell's theory, with or without matter, has another higher-form symmetry, associated with the Bianchi identity $\dd\deg F2=0$. The corresponding conserved current is proportional to $\ho\deg F2$. This is a topological $(D-3)$-form symmetry referred to as ``magnetic.'' To see why, one just needs to restrict to $D=4$ and observe that for a spatial surface $\S_2$, the charge~\eqref{charge} measures magnetic flux through $\S_2$. Intuitively, one can follow the analogy with the electric 1-form symmetry and conclude that the charged object now is the worldline of a magnetically charged particle (monopole) surrounded by the surface $\S_2$. In the language of gauge theory, this corresponds to a 't Hooft line. This example highlights yet another important difference between Noether and topological symmetries. Namely, since there are no local dynamical degrees of freedom carrying magnetic charge in Maxwell's theory, the magnetic monopole should not be viewed as a dynamical object, but rather as a static defect whose insertion modifies the Hilbert space of the theory.


\subsection{Coupling to background gauge fields}
\label{subsec:backgroungauge}

One practically useful way to probe the symmetry of a given field theory is to couple its conserved currents to a set of background gauge fields. In the context of spontaneously broken 0-form internal symmetries, this strategy has been developed into a versatile tool with far-reaching applications to particle phenomenology. The ensuing background gauge invariance of the generating functional provides among others an efficient way of constructing an effective action for NG bosons of the broken symmetry~\cite{Leutwyler1994a,Leutwyler1994b}. Here I will use it primarily to identify the higher-group symmetry in effective theories with composite currents.

A $p$-form symmetry is associated with a $(p+1)$-form current $\deg J{p+1}$ and thus in turn with a $(p+1)$-form background gauge field $\deg A{p+1}$. Noether symmetries can be gauged using standard techniques, barring possible obstructions such as 't Hooft anomalies. Once the gauged action, $S[A]$, is known, the gauged Noether currents can be identified from
\begin{equation}
\ho\deg J{p+1}\equiv\frac{\d S}{\d\deg A{p+1}},
\label{currentdef}
\end{equation}
where the variational derivative with respect to the gauge field is implicitly understood to act on the action from the left. In bosonic theories, the dependence of the action on the background gauge fields for Noether symmetries is typically nonlinear, and accordingly the current~\eqref{currentdef} is modified in the presence of gauge fields. Provided that all the symmetries of the given theory have been identified correctly, the divergence of the gauged currents should either vanish or be a local function of the background fields alone. In the latter case, one speaks of the 't Hooft anomaly. By construction, whenever the current defined by eq.~\eqref{currentdef} is anomalous, its divergence corresponds to the \emph{consistent} anomaly~\cite{Bertlmann1996a}.

In contrast to Noether currents, topological currents can be coupled to their respective gauge fields strictly linearly, by adding a new term to the action,\footnote{Throughout the paper, integration is implicitly assumed --- unless explicitly stated otherwise --- to be performed over the $D$-dimensional spacetime manifold, which is for the sake of simplicity assumed to have no boundary.}
\begin{equation}
S\supset\int\deg A{p+1}\w\ho\deg J{p+1}.
\label{AJcoupling}
\end{equation}
This suggests the approach I will take to construct effective actions for theories with a higher-group symmetry. Namely, I will assume that the Noether symmetries whose spontaneous breakdown dictates the dynamical degrees of freedom of the low-energy EFT have already been gauged. Then I will identify a set of trivially conserved currents and gauge them with respect to the Noether symmetries. In the last step, I will add a coupling of these gauged topological currents to their own background gauge fields via eq.~\eqref{AJcoupling}.


\section{Emergent topological currents from cohomology}
\label{sec:topcurrents}

As was stressed in the introduction, the mechanism leading to higher-group symmetry, proposed in this paper, relies essentially on the presence of topological currents in the given field theory. It is, in fact, not a problem to find systems with a trivially conserved current. Indeed, \emph{any} field theory that contains, among other degrees of freedom, a real scalar field $\p$ possesses such a current, namely $\ho\dd\p$. This guarantees a trivial local conservation law, but does not in itself guarantee the existence of a corresponding global symmetry. For the latter, the spectrum of the system should exhibit states of finite energy carrying nonzero integral charge~\eqref{charge}.

It was already observed in section~\ref{subsec:charges} that the integral charge can only be nonzero when the surface $\S_{D-p-1}$ belongs to a nontrivial homology class. By the same token, the Hodge-dual current $\ho\deg J{p+1}$ should belong to a nontrivial cohomology class of closed $(p+1)$-forms. Here the cohomology is not understood in terms of differential forms on the spacetime manifold, but rather in terms of differential forms on the jet space of the given field theory, that is, forms whose coefficients are local functions of the fields and a finite number of their derivatives. The ``characteristic cohomology'' of nontrivial forms on the jet space, closed modulo the equations of motion, was analyzed in refs.~\cite{Barnich1995a,Barnich2000a} using BRST techniques, assuming flat Euclidean spacetime and a star-shaped target space from which the fields take their values. However, in systems with a spontaneously broken symmetry, trivially conserved currents yielding excitations that carry topological charge arise, as a rule, from the nontrivial topology of the vacuum manifold. A comprehensive classification of such topological currents does not seem to be available in the literature. Here I would like to point out that there \emph{is} a class of topological currents that can be constructed by exploiting well-known mathematical results for the so-called Wess-Zumino (WZ) terms.

The possible presence of WZ terms in the action of the given field theory has attracted attention for decades, thanks to their relevance for particle phenomenology, close relation to anomalies and, last but not least, their intrinsic appeal. A geometric construction of gauged WZ terms in a scope and form relevant for strong nuclear interactions was advanced nearly four decades ago~\cite{Kaymakcalan1984a,Manohar1984b,Alvarez-Gaume1985a,Manes1985a}, soon after Witten elucidated their geometric nature~\cite{Witten1983a}. Quite recently, the intricate topological properties of WZ and other topological terms have become subject to a renewed scrutiny~\cite{Davighi2018a,Lee2020a,Yonekura2020a,Davighi2020a}. It has however been known already since the 1990s that in systems where a continuous global compact (0-form) symmetry group $G$ is spontaneously broken to its subgroup $H$, possible WZ terms in the low-energy EFT for the ensuing NG bosons are determined by de Rham cohomology of the coset space $G/H$, modulo some assumptions on the topology of the coset space and of the spacetime manifold~\cite{DHoker1994a,DHoker1995b}.

The cohomology generators, constructed explicitly in ref.~\cite{DHoker1995b}, can play multiple roles depending on the dimension of spacetime they are pulled back to. Namely, a given nontrivial closed $p$-form $\deg\om p$ on the coset space $G/H$ produces a WZ term when pulled back to a spacetime of dimension $D=p-1$. It can however also be used to construct a so-called $\theta$-term in $D=p$ dimensions, that is a surface term in the action that can only acquire a nonzero value for topologically nontrivial field configurations. Finally, in $D>p$ dimensions, the cohomology generator $\deg\om p$ can be interpreted as the Hodge dual of a topological $(D-p)$-form current. As such, it gives rise to a $(D-p-1)$-form symmetry. The various physical interpretations of cohomology generators with $p\leq5$ --- which are those relevant in $D\leq4$ spacetime dimensions --- are summarized for the reader's convenience in table~\ref{tab:WZforms}.

\begin{table}
\begin{center}
\renewcommand{\arraystretch}{1.2}
\begin{tabular}{c|ccccc}
\toprule
& $\vek{p=1}$ & $\vek{p=2}$ & $\vek{p=3}$ & $\vek{p=4}$ & $\vek{p=5}$\\
\midrule
$\vek{D=1}$ & $\theta$-term & WZ term & --- & --- & ---\\
$\vek{D=2}$ & 1-form current & $\theta$-term & WZ term & --- & ---\\
$\vek{D=3}$ & 2-form current & 1-form current & $\theta$-term & WZ term & ---\\
$\vek{D=4}$ & 3-form current & 2-form current & 1-form current & $\theta$-term & WZ term\\
\bottomrule
\end{tabular}
\end{center}
\caption{Physical interpretation of $p$-form cohomology generators depending on the spacetime dimension $D$. Only generators relevant in $D\leq4$ dimensions are included. In general, a cohomology generator of degree $p$ on the coset space $G/H$ pulled back to $D>p$ spacetime dimensions gives a topologically conserved $(D-p)$-form current, and hence a $(D-p-1)$-form symmetry.}
\label{tab:WZforms}
\end{table}

At first sight, it looks like we have actually managed to classify topological currents, at least in the context of low-energy EFTs for NG bosons. The above observation should, however, be taken with a grain of salt. All the cohomology generators responsible for a WZ term in the action are by construction invariant under the full symmetry group $G$. Hence, they give a likewise $G$-invariant topological current in spacetimes of higher dimension. This is obviously not the most general possibility, and one can imagine currents that transform in some nontrivial representation of the global symmetry. To classify such topological currents, or even find a set of physically relevant examples, remains an interesting open question. On the other hand, the good news is that the usual homotopy constraints, which are required for the consistency of the higher-dimensional formulation of WZ terms yet have drawn some criticism~\cite{Davighi2018a}, are now absent. Once the $p$-form cohomology generator~$\deg\om p$ is interpreted as a $(D-p)$-form current in $D$ dimensions, there is no need to impose a priori constraints on the topology of the coset space $G/H$ or the spacetime.


\subsection{Explicit construction of topological currents}
\label{subsec:DHoker}

I will now review some explicit results for cohomology generators on a coset space $G/H$. I will only include cohomology generators with $p\leq3$, which are those that produce topological currents in $D\leq4$ spacetime dimensions; explicit expressions for cohomology generators up to $p=5$ are given in refs.~\cite{DHoker1995b,Brauner2019a}. I will at first assume that $G$ is compact, semisimple and simply connected, and that $H$ is connected.

Some further notation will be needed. The generators of $G$ will be denoted as $T_{i,j,\dotsc}$. In order to distinguish broken and unbroken generators, the Latin indices $a,b,\dotsc$ will be used for the former and the Greek indices $\a,\b,\dotsc$ for the latter. The Lie algebra of $G$ is then fixed by the structure constants $f_{ij}^{k}$ such that
\begin{equation}
[T_i,T_j]=f_{ij}^{k}T_k.
\end{equation}
A set of background gauge fields for the (0-form) symmetry group $G$ is encoded in the 1-form connection $\deg A1$, taking values in the Lie algebra of $G$. Under a gauge transformation with coordinate-dependent parameter $g\in G$, it transforms as usual as
\begin{equation}
\deg A1\xrightarrow{g}g\deg A1g^{-1}+g\dd g^{-1},
\label{Atransfo}
\end{equation}
with the corresponding $G$-covariant field strength 2-form being
\begin{equation}
\deg F2\equiv\dd\deg A1+\deg A1\w\deg A1.
\end{equation}
The NG boson degrees of freedom of the coset space $G/H$ are introduced via a matrix-valued field $U$. It is customary to think of this as $U=e^{-\pi^aT_a}$, where $\pi^a$ are the individual NG fields, but this amounts to a specific choice of coordinates on the coset space and will not be necessary. All that matters is that $U$ parametrizes the coset space $G/H$ in a chosen faithful representation of $G$.

The fundamental object, in terms of which the cohomology generators are constructed, is the Lie-algebra-valued (gauged) Maurer-Cartan (MC) 1-form,
\begin{equation}
\deg\theta1\equiv U^{-1}(\dd+\deg A1)U\equiv\deg\theta1{}^iT_i.
\label{MCform}
\end{equation}
As is well-known from the coset construction of effective Lagrangians~\cite{Coleman1969a,Callan1969a}, the broken part of the MC form, $\deg\p1\equiv\deg\theta1{}^aT_a$, transforms covariantly under the whole symmetry group $G$ and constitutes the basic building block for the construction of $G$-invariant operators. The unbroken part of the MC form, $\deg V1\equiv\deg\theta1{}^\a T_\a$, transforms as a 1-form connection taking values in the Lie algebra of $H$. It gives rise to the field strength 2-form
\begin{equation}
\deg W2\equiv\dd\deg V1+\deg V1\w\deg V1\equiv\deg W2{}^\a T_\a.
\end{equation}
A similar field strength 2-form can be composed out of the whole MC form $\deg\theta1$. This leads to the so-called MC structure equation,
\begin{equation}
\dd\deg\theta1+\deg\theta1\w\deg\theta1=U^{-1}\deg F2U\equiv\deg{\bar F}2\equiv\deg{\bar F}2{}^iT_i.
\label{MCeq}
\end{equation}

We now have all the pieces needed to spell out the results for cohomology generators of degree $p\leq3$. In fact, with the assumptions that $G$ is simply connected and $H$ is connected, it follows at once that the coset space $G/H$ is simply connected (see for instance the overview of homotopy groups in physics in ref.~\cite{Abanov2017a}). The Hurewicz theorem~\cite{Frankel2012a} then in turn implies that the first cohomology group of $G/H$ is trivial. Cohomology generators of degree 2 take the general form
\begin{equation}
\deg\Om2=-c_\a\deg W2{}^\a=\frac12c_\a f^\a_{bc}\deg\theta1{}^b\w\deg\theta1{}^c-c_\a\deg{\bar F}2{}^\a.
\label{cohom2}
\end{equation}
Here the set of constants $c_\a$ is required to satisfy the $H$-invariance condition
\begin{equation}
c_\a f^\a_{\b\g}=0,
\end{equation}
which implies that for compact Lie groups, degree-2 cohomology generators on the coset space $G/H$ are in a one-to-one correspondence with $\gr{U(1)}$ factors of $H$.

In case of $p=3$, the cohomology generators are parametrized by constant symmetric $G$-invariant tensors $d_{ij}$ that vanish on the unbroken subgroup, that is $d_{\a\b}=0$. They are most easily expressed in a matrix form by noting that for compact semisimple groups $G$, the most general invariant tensor $d_{ij}$ reads
\begin{equation}
d_{ij}=\sum_\sigma d_\sigma\tr_\sigma(T_iT_j),
\end{equation}
where the index $\sigma$ labels different simple factors of $G$ and the trace $\tr_\sigma$ is done over the $\sigma$-th simple component. The possible values of the coefficients $d_\sigma$ are further restricted by the requirement that $d_{\a\b}=0$. The corresponding cohomology generator of degree 3 then takes the form
\begin{equation}
\deg\Om3=\sum_\sigma d_\sigma\tr_\sigma\left[\frac13\deg\p1\w\deg\p1\w\deg\p1-(\deg W2+\deg{\bar F}2)\w\deg\p1\right].
\label{cohom3}
\end{equation}
As a matter of fact, this 3-form is only closed in the absence of background gauge fields, or when only gauge fields for the unbroken subgroup $H$ are present. Background fields for the broken generators present an obstruction to gauging the form, closely related to the chiral anomaly in $D=2$ spacetime dimensions and expressed by the relation
\begin{equation}
\dd\deg\Om3=-\sum_\sigma d_\sigma\tr_\sigma(\deg F2\w\deg F2).
\end{equation}
A similar obstruction appears for all cohomology generators of odd degrees. In the interpretation of cohomology generators in terms of topological currents, it signals the impossibility to construct a current that is simultaneously conserved and gauge-invariant. When the topological current is coupled to a new background gauge field of its own, the obstruction manifests itself by a mixed 't Hooft anomaly in the generating functional of the theory.

Let us now see what happens when the coset space $G/H$ is not simply connected~\cite{DHoker1995b}. This may happen when $G$ is not simply connected or $H$ is not connected. The first cohomology group of $G/H$ may then contain nontrivial generators $\deg\Om1$; the simplest example is fully broken $\gr{U(1)}$ symmetry with the de Rham cohomology group $H^1_\text{dR}(G/H)\cong H^1_\text{dR}(S^1)\cong\R$. In general, a $G$-covariant 1-form assumes the form
\begin{equation}
\deg\Om1=e_a\deg\theta1{}^a.
\label{cohom1}
\end{equation}
The conditions of $G$-invariance and closedness require the coefficient $e_a$ to satisfy $e_af^a_{ij}=0$. For compact symmetry groups, only $\gr{U(1)}$ factors of $G$ can therefore give rise to degree-1 cohomology generators. It follows from eq.~\eqref{MCeq} that
\begin{equation}
\dd\deg\Om1=e_a\deg F2{}^a,
\end{equation}
implying the same type of anomaly as for the 3-form~\eqref{cohom3}.

With the degree-1 generators at hand, it is now possible to form additional generators of higher cohomology groups through products of lower-degree generators; generators that can or cannot be decomposed into (linear combinations of) exterior products of lower-degree generators are referred to respectively as ``decomposable'' and ``primitive''~\cite{DHoker1995b}. The structure of possible generators of cohomology groups with $p\leq3$ is reviewed in table~\ref{tab:decomposable}.

\begin{table}
\begin{center}
\renewcommand{\arraystretch}{1.2}
\begin{tabular}{c|cc}
\toprule
& \textbf{primitive} & \textbf{decomposable}\\
\midrule
$\vek{p=1}$ & $\deg\Om1$ & ---\\
$\vek{p=2}$ & $\deg\Om2$ & $\deg\Om1\w\deg\Om1$\\
$\vek{p=3}$ & $\deg\Om3$ & $\deg\Om1\w\deg\Om1\w\deg\Om1$, $\deg\Om1\w\deg\Om2$\\
\bottomrule
\end{tabular}
\end{center}
\caption{Schematic list of possibilities how to construct cohomology generators of a given degree. In general, the list of possible decomposable cohomology generators is determined by the partitions of the cohomology degree: degree-$p$ cohomology includes all decomposable generators of the form $\deg\Om{p_1}\w\deg\Om{p_2}\w\dotsb$ such that $1\leq p_1\leq p_2\leq\dotsb$ and $p_1+p_2+\dotsb=p$.}
\label{tab:decomposable}
\end{table}

To summarize the contents of this section, theories with a spontaneously broken compact 0-form symmetry group $G$ may possess $G$-invariant topological currents, classified by the de Rham cohomology of the coset space $G/H$. When pulled back to a spacetime of dimension $D>p$, a cohomology generator of degree $p$ gives rise to a topologically conserved $(D-p)$-form current, and accordingly a $(D-p-1)$-form symmetry. Cohomology generators of degree 1 typically arise from $\gr{U(1)}$ factors of the symmetry group. Primitive generators of degree 2 and 3 are given respectively by eqs.~\eqref{cohom2} and~\eqref{cohom3}. Additional, decomposable generators of higher degrees may be constructed as shown in table~\ref{tab:decomposable}.

Out of the topological currents, included explicitly in table~\ref{tab:WZforms}, those in the $p=1$ column are featured in Abelian superfluids, as already observed in ref.~\cite{Gaiotto2015a}. The case of $p=2$ and $D=3$ or $D=4$, giving rise respectively to a 0-form or a 1-form symmetry, is relevant for (anti)ferromagnets, where $G/H=\gr{SU(2)}/\gr{U(1)}$ so that $H^2_\text{dR}(G/H)\cong H^2_\text{dR}(S^2)\cong\R$. The corresponding 1-form or 2-form current counts (anti)ferromagnetic skyrmions. This is a special case of the general class of $\CP^N$ models, mentioned in ref.~\cite{Cordova2019a}. Finally, the 1-form current corresponding to $p=3$ and $D=4$ appears for instance as the skyrmion number current in the low-energy EFT of quantum chromodynamics (QCD), as pointed out in ref.~\cite{Delacretaz2020a}, and is seen via anomaly matching to correspond to baryon number.


\section{Abelian theories with second-order composite currents}
\label{sec:abelian}

We are now ready to start analyzing the symmetries of concrete field theories possessing topologically conserved composite currents. The minimal example of such a theory would be one with two degrees of freedom and a single second-order composite current. It comes, however, at a little extra cost to work out the details for the much broader class of theories with an arbitrary number of constituents, the only limitation being to pairwise constructed second-order composite currents.

Consider a set of $p_i$-form Abelian symmetries distinguished by the flavor index $i$, spontaneously broken so that the low-energy description of the system is an EFT of the corresponding $p_i$-form NG fields, $\deg{\p_i}{p_i}$. Associated with these is a set of $(p_i+1)$-form background gauge fields, $\deg{A_i}{p_i+1}$. In the current presentation, I \emph{choose} to make the invariance under these given, Noether-type Abelian symmetries manifest. This can be guaranteed by making sure that the derivatives $\dd\deg{\p_i}{p_i}$ only enter the action through the covariant exterior derivatives, $\deg{\Om_i}{p_i+1}=\dd\deg{\p_i}{p_i}-\deg{A_i}{p_i+1}$. Other possible choices will be discussed in section~\ref{subsec:ambiguities}.

As explained at length in the preceding sections, the presence of the NG fields induces a set of emergent $(D-p_i-2)$-form symmetries, associated to the topological currents, defined up to an arbitrary normalization factor as $\ho\deg{\Om_i}{p_i+1}$. These emergent symmetries can be probed by adding a set of $(D-p_i-1)$-form background gauge fields, $\codeg{B_i}{p_i+1}$. In addition, one can construct second-order topological currents, defined again up to normalization as $\ho\bigl(\deg{\Om_i}{p_i+1}\w\deg{\Om_j}{p_j+1}\bigr)$. These are coupled to a set of $(D-p_i-p_j-2)$-form background gauge fields, $\codeg{C_{ij}}{p_i+p_j+2}$. By the graded symmetry of the exterior product, the latter must satisfy
\begin{equation}
\codeg{C_{ji}}{p_i+p_j+2}=(-1)^{(p_i+1)(p_j+1)}\codeg{C_{ij}}{p_i+p_j+2}.
\label{Csym}
\end{equation}

A generic action featuring the NG fields as well as all the background fields just introduced then takes the form\footnote{Here and in the following, sums over flavor indices $i,j$ are explicitly indicated, unless the same index appears in a given operator exactly twice so that the Einstein summation convention can be used.}
\begin{equation}
\begin{split}
S=S_\text{inv}+\int\biggl[&\k_1\sum_i\codeg{B_i}{p_i+1}\w(\dd\deg{\p_i}{p_i}-\deg{A_i}{p_i+1})\\
&+\frac{\k_2}2\sum_{i,j}\codeg{C_{ij}}{p_i+p_j+2}\w(\dd\deg{\p_i}{p_i}-\deg{A_i}{p_i+1})\w(\dd\deg{\p_j}{p_j}-\deg{A_j}{p_j+1})\biggr],
\end{split}
\label{masteraction_Abelian}
\end{equation}
where $\k_{1,2}$ are couplings that fix the normalization of the source terms, and $S_\text{inv}$ stands for a part of the action that is manifestly gauge-invariant, being constructed solely out of the covariant exterior derivatives of $\deg{\p_i}{p_i}$ and independent of the background fields $\codeg{B_i}{p_i+1}$ and $\codeg{C_{ij}}{p_i+p_j+2}$. This part of the action by default includes the kinetic term, proportional to $\sum_i\deg{\Om_i}{p_i+1}\w\ho\deg{\Om_i}{p_i+1}$, as well as higher-order interaction terms. In superfluids, it is known to be fixed by the equation of state of the system~\cite{Greiter1989a,Son2002a}.

The whole master action~\eqref{masteraction_Abelian} is by construction manifestly invariant under the combined gauge transformation of the NG fields $\deg{\p_i}{p_i}$ and the background fields $\deg{A_i}{p_i+1}$,
\begin{equation}
\deg{\p_i}{p_i}\to\deg{\p_i}{p_i}+\deg{\a_i}{p_i},\qquad
\deg{A_i}{p_i+1}\to\deg{A_i}{p_i+1}+\dd\deg{\a_i}{p_i},
\label{Noethersym}
\end{equation}
where $\deg{\a_i}{p_i}$ is a $p_i$-form symmetry parameter. Due to the Abelian nature of all the symmetries involved, we might furthermore guess the analogous transformations for the background fields $\codeg{B_i}{p_i+1}$ and $\codeg{C_{ij}}{p_i+p_j+2}$,
\begin{equation}
\codeg{B_i}{p_i+1}\to\codeg{B_i}{p_i+1}+\dd\codeg{\b_i}{p_i+2},\quad
\codeg{C_{ij}}{p_i+p_j+2}\to\codeg{C_{ij}}{p_i+p_j+2}+\dd\codeg{\g_{ij}}{p_i+p_j+3}\qquad
\text{(wrong)}.
\label{wrong}
\end{equation}
This works for the $\k_1$ term in the action, leaving behind a mixed 't Hooft anomaly for the $\deg{A_i}{p_i+1}$ and $\codeg{B_i}{p_i+1}$ fields. On the contrary, it does not work for the $\k_2$ term due to the fact that upon gauging, the second-order composite current that $\codeg{C_{ij}}{p_i+p_j+2}$ couples to is no longer conserved, and moreover its divergence depends nontrivially on the dynamical fields $\deg{\p_i}{p_i}$. This problem can, however, be fixed by modifying the transformation rule for $\codeg{B_i}{p_i+1}$. Namely, it is easy to verify that the undesired $\deg{\p_i}{p_i}$-dependent variation of the action is removed if all the fields undergo the following simultaneous set of transformations,
\begin{equation}
\begin{split}
\deg{\p_i}{p_i}&\to\deg{\p_i}{p_i}+\deg{\a_i}{p_i},\\
\deg{A_i}{p_i+1}&\to\deg{A_i}{p_i+1}+\dd\deg{\a_i}{p_i},\\
\codeg{B_i}{p_i+1}&\to\codeg{B_i}{p_i+1}+\dd\codeg{\b_i}{p_i+2}-\frac{\k_2}{\k_1}\sum_j(-1)^{D+p_ip_j}\codeg{\g_{ij}}{p_i+p_j+3}\w\dd\deg{A_j}{p_j+1},\\
\codeg{C_{ij}}{p_i+p_j+2}&\to\codeg{C_{ij}}{p_i+p_j+2}+\dd\codeg{\g_{ij}}{p_i+p_j+3}.
\end{split}
\label{gauge_Abelian}
\end{equation}
The variation of the action then becomes, apart from surface terms that will generally be discarded throughout the paper,\footnote{\label{ftn:surface}All surface terms that appear in the analysis of the background symmetries of the action throughout the paper are integrals of exact forms, which vanish on spacetimes without boundary. This can be ensured by a suitable restriction of the setup, notably by only including background gauge fields with a globally well-defined connection, and assuming likewise that the gauge transformation parameters $\deg{\a_i}{p_i}$, $\codeg{\b_i}{p_i+2}$ and $\codeg{\g_{ij}}{p_i+p_j+3}$ are globally well-defined. Extending the background gauge invariance beyond these limitations may result in quantization of the couplings $\k_{1,2}$, depending on the concrete choice of symmetry group.}
\begin{equation}
\d S=\int\sum_i(-1)^{D+p_i}\k_1\codeg{\b_i}{p_i+2}\w\dd\deg{A_i}{p_i+1}.
\label{dS_Abelian}
\end{equation}
This corresponds to a mixed 't Hooft anomaly, as will be explained in detail in section~\ref{subsec:ambiguities}.

Let us pause and think about what we have found. In the absence of the background fields $\deg{A_i}{p_i+1}$, the action~\eqref{masteraction_Abelian} has two independent, naive higher-form symmetries as indicated by eq.~\eqref{wrong}. Should we choose to erase the composite current and the associated background field $\codeg{C_{ij}}{p_i+p_j+2}$, we would still find a good symmetry under independent transformations of $\deg{A_i}{p_i+1}$ and $\codeg{B_i}{p_i+1}$, albeit with an obstruction encoded in the 't Hooft anomaly~\eqref{dS_Abelian}. To get a firm grasp on the symmetry of the theory, it is however mandatory to consider simultaneously sources for \emph{all} of its conserved currents. We find that doing so requires a deformation of the transformation laws of the background fields as in eq.~\eqref{gauge_Abelian}.

In the physically simplest case where $p_i=0$ for all the Noether symmetries and one restricts to $D=3$ spacetime dimensions, the $\deg{A_i}1$ and $\codeg{C_{ij}}2$ fields couple to 0-form symmetries, whereas the $\codeg{B_i}1$ fields couple to 1-form symmetries. The transformation rules~\eqref{gauge_Abelian} then precisely match the 2-group structure detailed by C\'ordova et al.~\cite{Cordova2019a}, where the transformation of the 2-form background field picks a contribution proportional to the transformation parameter of a 0-form symmetry and the field strength of a 1-form background gauge field. The pattern is however obviously more general and is not restricted to a particular choice of $p_i$ and spacetime dimension.

Modification of the transformation rules for the background fields requires a corresponding modification of the field strengths. This is straightforward to find,
\begin{equation}
\begin{split}
\deg{F_i}{p_i+2}&=\dd\deg{A_i}{p_i+1},\\
\codeg{G_i}{p_i}&=\dd\codeg{B_i}{p_i+1}+\frac{\k_2}{\k_1}\sum_j(-1)^{D+p_ip_j}\codeg{C_{ij}}{p_i+p_j+2}\w\dd\deg{A_j}{p_j+1},\\
\codeg{H_{ij}}{p_i+p_j+1}&=\dd\codeg{C_{ij}}{p_i+p_j+2}.
\end{split}
\label{FGH_Abelian}
\end{equation}
All the three field strength forms are now invariant under the simultaneous transformation of the gauge fields~\eqref{gauge_Abelian}. Modification of the field strengths in turn affects the Bianchi identities that they satisfy,
\begin{equation}
\dd\deg{F_i}{p_i+2}=0,\quad
\dd\codeg{G_i}{p_i}=\frac{\k_2}{\k_1}\sum_j(-1)^{D+p_ip_j}\codeg{H_{ij}}{p_i+p_j+1}\w\deg{F_j}{p_j+2},\quad
\dd\codeg{H_{ij}}{p_i+p_j+1}=0.
\label{Bianchi_Abelian}
\end{equation}


\subsection{Current conservation and Ward identities}
\label{subsec:ward}

While the background field invariance of the action or the generating functional is a practically useful tool to identify the symmetry of the theory, it is the current conservation laws and the ensuing Ward identities that relate directly to the correlation functions of the theory, and thus observables. Following the general definition~\eqref{currentdef} of a current through a variation of the action with respect to the associated background field, we can set
\begin{equation}
\ho\deg{J_{Ai}}{p_i+1}\equiv\frac{\d S}{\d\deg{A_i}{p_i+1}},\qquad
\ho\codeg{J_{Bi}}{p_i+1}\equiv\frac{\d S}{\d\codeg{B_i}{p_i+1}},\qquad
\ho\codeg{J_{Cij}}{p_i+p_j+2}\equiv\frac{\d S}{\d\codeg{C_{ij}}{p_i+p_j+2}}.
\label{JABCdef}
\end{equation}
The variation of action~\eqref{dS_Abelian} under the gauge transformations~\eqref{gauge_Abelian} then implies the set of Ward identities expressed as closure relations for the Hodge-dual currents,\footnote{\label{ftn:diagonal}Here I am being a bit cavalier toward the double sum over $i,j$, involved in the $\k_2$ term of the action. In the cases where diagonal, $i=j$ contributions are present, which is compatible with the graded symmetry of exterior product when $p_i$ is odd, the right-hand side of the relation for $\dd\ho\codeg{J_{Cii}}{2p_i+2}$ will acquire an extra factor of $1/2$ compared to eq.~\eqref{Ward_Abelian}. Likewise, the current $\ho\codeg{J_{Cii}}{2p_i+2}$ implied by eq.~\eqref{JBC} should be corrected by a factor of $1/2$.}
\begin{equation}
\begin{split}
\dd\ho\deg{J_{Ai}}{p_i+1}&=0,\\
\dd\ho\codeg{J_{Bi}}{p_i+1}&=-\k_1\dd\deg{A_i}{p_1+1},\\
\dd\ho\codeg{J_{Cij}}{p_i+p_j+2}&=-\frac{\k_2}{\k_1}\Bigl[\dd\deg{A_i}{p_i+1}\w\ho\codeg{J_{Bj}}{p_j+1}+(-1)^{(p_i+1)(p_j+1)}\dd\deg{A_j}{p_j+1}\w\ho\codeg{J_{Bi}}{p_i+1}\Bigr].
\end{split}
\label{Ward_Abelian}
\end{equation}
While the precise form of the Noether current $\deg{J_{Ai}}{p_i+1}$ depends sensitively on the dynamics of the system through the invariant part of the action $S_\text{inv}$, the topological currents $\codeg{J_{Bi}}{p_i+1}$ and $\codeg{J_{Cij}}{p_i+p_j+2}$ are determined completely by the source terms in the action~\eqref{masteraction_Abelian} and read
\begin{equation}
\begin{split}
\ho\codeg{J_{Bi}}{p_i+1}&=\k_1(\dd\deg{\p_i}{p_i}-\deg{A_i}{p_i+1}),\\
\ho\codeg{J_{Cij}}{p_i+p_j+2}&=\k_2(\dd\deg{\p_i}{p_i}-\deg{A_i}{p_i+1})\w(\dd\deg{\p_j}{p_j}-\deg{A_j}{p_j+1}).
\end{split}
\label{JBC}
\end{equation}
The corresponding topological conservation laws can thus easily be verified explicitly.

The relation for $\dd\ho\codeg{J_{Bi}}{p_i+1}$ encodes the mixed 't Hooft anomaly for the $p_i$-form Noether symmetry and its $(D-p_i-2)$-form dual. It was shown in ref.~\cite{Delacretaz2020a} that this anomaly can be taken as a starting point of a proof of existence of a massless NG boson, alternative to the usual Goldstone theorem based on the assumption of the existence of long-range order. The fact that the anomaly appears in the $\codeg{J_{Bi}}{p_i+1}$ current is due to the choice I made to keep the invariance under the gauge transformation of $\deg{A_i}{p_i+1}$ manifest. This choice can be altered by adding a local functional of the background gauge fields to the action, as I will explain shortly.

The Ward identity for $\codeg{J_{Cij}}{p_i+p_j+2}$ is of the fusion type discussed in ref.~\cite{Cordova2019a}, and it is one of the hallmarks of a higher-group symmetry. Equation~\eqref{Ward_Abelian} together with the transformation rules~\eqref{gauge_Abelian} constitutes one of the main results of this paper, showing that a higher-group symmetry structure is an inevitable consequence of the existence of a composite topological current in the given theory. The remainder of the paper is devoted to an elaboration of this result. In the following two subsections, I will discuss the detailed properties of the mixed 't Hooft anomaly already mentioned several times, and give a couple of concrete examples of physical systems featuring the higher-group symmetry structure put forward here. The next three sections then introduce various generalizations of the basic result just presented, as well as a dual picture where the roles of the Noether and primitive topological symmetries are interchanged.


\subsection{Ambiguities in the anomaly}
\label{subsec:ambiguities}

The variation of the action under the combined gauge transformations~\eqref{gauge_Abelian} can be fully described by a local function of the background gauge fields and the transformation parameters, $\deg\I D$ in a notation borrowed from ref.~\cite{Cordova2019a}, defined by
\begin{equation}
\d S=2\pi\im\int\deg\I D.
\label{IDdef}
\end{equation}
It follows from eq.~\eqref{dS_Abelian} that for the choice of action~\eqref{masteraction_Abelian}, one has
\begin{equation}
\deg\I D=\frac{\k_1}{2\pi\im}\sum_i(-1)^{D+p_i}\codeg{\b_i}{p_i+2}\w\dd\deg{A_i}{p_i+1}.
\label{IDAbelian}
\end{equation}
If we imagine that the $D$-dimensional spacetime manifold forms the boundary of a $(D+1)$-dimensional bulk, this can be understood via the anomaly inflow mechanism as arising from a Chern-Simons (CS) theory in the bulk. The action of the CS theory is defined by local function $\deg\I{D+1}$ of the gauge fields such that its variation under the transformations~\eqref{gauge_Abelian} satisfies $\d\deg\I{D+1}\equiv\dd\deg\I D$. In our case, it readily follows that
\begin{equation}
\deg\I{D+1}=\frac{\k_1}{2\pi\im}\sum_i(-1)^{D+p_i}\codeg{B_i}{p_i+1}\w\dd\deg{A_i}{p_i+1}.
\label{ID1Abelian}
\end{equation}
Note that the last, nontrivial contribution to the transformation of $\codeg{B_i}{p_i+1}$ in eq.~\eqref{gauge_Abelian} drops out of $\d\deg\I{D+1}$ thanks to the graded symmetry of $\codeg{\g_{ij}}{p_i+p_j+3}$, inherited from that of $\codeg{C_{ij}}{p_i+p_j+2}$ via eq.~\eqref{Csym}. Finally, the exterior derivative of the CS $(D+1)$-form~\eqref{ID1Abelian} gives a gauge-invariant anomaly polynomial, $\deg\I{D+2}\equiv\dd\deg\I{D+1}$, that governs the 't Hooft anomalies of the theory. Using the same graded symmetry of $\codeg{C_{ij}}{p_i+p_j+2}$, this can be expressed directly in terms of the field strengths~\eqref{FGH_Abelian},
\begin{equation}
\deg\I{D+2}=\frac{\k_1}{2\pi\im}\sum_i(-1)^{D+p_i}\dd\codeg{B_i}{p_i+1}\w\dd\deg{A_i}{p_i+1}=\frac{\k_1}{2\pi\im}\sum_i(-1)^{D+p_i}\codeg{G_i}{p_i}\w\deg{F_i}{p_i+2}.
\label{ID2Abelian}
\end{equation}

It is well-known that unlike the gauge-invariant anomaly polynomial~\eqref{ID2Abelian}, the $D$-dimensional description of the anomaly encoded in the $D$-form $\deg\I D$ defined by eq.~\eqref{IDdef} suffers from an ambiguity. Indeed, one can add to the action an arbitrary local functional of the background gauge fields only, $2\pi\im\int\deg\K D$, that serves as a counterterm to modify the current conservation laws~\eqref{Ward_Abelian}. Adding such a local counterterm amounts to shifting $\deg\I D$ by $\deg{\I_\text{c.t.}}D\equiv\d\deg\K D$ and in turn, via the descent relation $\d\deg\I{D+1}=\dd\deg\I D$, to shifting the CS $(D+1)$-form~\eqref{ID1Abelian} by $\deg{\I_\text{c.t.}}{D+1}=\dd\deg\K D$. This makes it clear that adding an arbitrary local functional of the background gauge fields to the action leaves the anomaly polynomial~\eqref{ID2Abelian} unchanged as expected.

While the counterterm functional can in principle be chosen at will, there are some natural options that suggest themselves. These are motivated by the observation that, by eqs.~\eqref{Ward_Abelian} and~\eqref{JBC}, the currents $\codeg{J_{Bi}}{p_i+1}$ and $\codeg{J_{Cij}}{p_i+p_j+2}$ are manifestly gauge-invariant but not conserved. On the other hand, the current $\deg{J_{Ai}}{p_i+1}$, while conserved, is not gauge-invariant due to contributions arising from the $\deg{A_i}{p_i+1}$-dependence of the source terms in eq.~\eqref{masteraction_Abelian}, that is due to the very gauge invariance of the other two currents. This is the usual dilemma in presence of an anomaly. For various reasons, one may want to make one or both of $\codeg{J_{Bi}}{p_i+1}$ and $\codeg{J_{Cij}}{p_i+p_j+2}$ conserved instead of $\deg{J_{Ai}}{p_i+1}$. This is a necessity should some of the gauge fields actually be dynamical.


\subsubsection{\boldmath Making $\codeg{J_{Bi}}{p_i+1}$ conserved}

The simplest choice of counterterm that suggests itself is
\begin{equation}
\deg\K D=\frac{\k_1}{2\pi\im}\sum_i\codeg{B_i}{p_i+1}\w\deg{A_i}{p_i+1}.
\label{ct1}
\end{equation}
This cancels the original variation of action as expressed by eq.~\eqref{IDAbelian} and replaces it, up to surface terms, with
\begin{equation}
\begin{split}
\deg\I D+\deg{\I_\text{c.t.}}D=\frac1{2\pi\im}\biggl[&\k_1\sum_i\codeg{B_i}{p_i+1}\w\dd\deg{\a_i}{p_i}\\
&-\k_2\sum_{i,j}(-1)^{D+p_j}\codeg{\g_{ij}}{p_i+p_j+3}\w(\deg{A_i}{p_i+1}+\dd\deg{\a_i}{p_i})\w\dd\deg{A_j}{p_j+1}\biggr].
\end{split}
\label{IDalt}
\end{equation}
This somewhat unwieldy expression corresponds to the inflow from a very simple bulk CS action, which is most easily obtained by adding $\dd\deg\K D$ to eq.~\eqref{ID1Abelian},
\begin{equation}
\deg{\I}{D+1}+\deg{\I_\text{c.t.}}{D+1}=\frac{\k_1}{2\pi\im}\sum_i\dd\codeg{B_i}{p_i+1}\w\deg{A_i}{p_i+1}.
\end{equation}
This of course yields the same anomaly polynomial~\eqref{ID2Abelian}. The variation of the action~\eqref{IDalt} can be used to obtain the modified Ward identities for the currents, still defined by eq.~\eqref{JABCdef}. A short calculation shows that the divergences of $\deg{J_{Ai}}{p_i+1}$ and $\codeg{J_{Bi}}{p_i+1}$ displayed in eq.~\eqref{Ward_Abelian} must be replaced with
\begin{equation}
\dd\ho\deg{J_{Ai}}{p_i+1}\to(-1)^{(D+1)(p_i+1)}\k_1\dd\codeg{B_i}{p_i+1},\qquad
\dd\ho\codeg{J_{Bi}}{p_i+1}\to0.
\end{equation}
The conservation of $\codeg{J_{Bi}}{p_i+1}$ is a simple consequence of the fact that adding the counterterm~\eqref{ct1} shifts the source terms in the action~\eqref{masteraction_Abelian} so that $\ho\codeg{J_{Bi}}{p_i+1}\to\k_1\dd\deg{\p_i}{p_i}$. Finally, the Ward identity for $\codeg{J_{Cij}}{p_i+p_j+2}$ keeps the same functional form as displayed in eq.~\eqref{Ward_Abelian}, except that the currents $\codeg{J_{Bi}}{p_i+1}$ on its right-hand side are to be interpreted as the  original ones, given by eq.~\eqref{JBC}. Altogether, the net effect of adding the counterterm~\eqref{ct1} to the action is to toggle which of the currents $\deg{J_{Ai}}{p_i+1}$ and $\codeg{J_{Bi}}{p_i+1}$ is conserved. This can be understood in terms of the duality between the associated symmetries, inspected more closely in section~\ref{sec:duality}.


\subsubsection{\boldmath Making $\codeg{J_{Bi}}{p_i+1}$ and $\codeg{J_{Cij}}{p_i+p_j+2}$ conserved}

What the counterterm~\eqref{ct1} does is to remove from the action~\eqref{masteraction_Abelian} the term proportional to $\codeg{B_i}{p_i+1}$ and independent of the dynamical NG fields $\deg{\p_i}{p_i}$. One may wish to do the same with the $\deg{\p_i}{p_i}$-independent term proportional to $\codeg{C_{ij}}{p_i+p_j+2}$ as well. To that end, we set
\begin{equation}
\deg\K D=\frac1{2\pi\im}\biggl(\k_1\sum_i\codeg{B_i}{p_i+1}\w\deg{A_i}{p_i+1}-\frac{\k_2}2\sum_{i,j}\codeg{C_{ij}}{p_i+p_j+2}\w\deg{A_i}{p_i+1}\w\deg{A_j}{p_j+1}\biggr)
\label{ct2}
\end{equation}
instead of eq.~\eqref{ct1}. The action can then be written in a compact manner as
\begin{equation}
\begin{split}
S+2\pi\im\int\deg\K D=S_\text{inv}+\int\biggl(&\k_1\sum_i\codeg{B_{\text{eff},i}}{p_i+1}\w\dd\deg{\p_i}{p_i}\\
&+\frac{\k_2}2\sum_{i,j}\codeg{C_{ij}}{p_i+p_j+2}\w\dd\deg{\p_i}{p_i}\w\dd\deg{\p_j}{p_j}\biggr),
\end{split}
\label{effaction}
\end{equation}
where
\begin{equation}
\codeg{B_{\text{eff},i}}{p_i+1}\equiv\codeg{B_i}{p_i+1}-\frac{\k_2}{\k_1}\sum_j\codeg{C_{ji}}{p_i+p_j+2}\w\deg{A_j}{p_j+1}.
\label{Beff}
\end{equation}
One may of course think of this as a mere shorthand notation and carry on with the analysis of the symmetries of the theory under the transformation~\eqref{gauge_Abelian} of the original background fields $\codeg{B_i}{p_i+1}$ and $\codeg{C_{ij}}{p_i+p_j+2}$. The variation of the action is then described, up to surface terms, by
\begin{equation}
\deg\I D+\deg{\I_\text{c.t.}}D=\frac1{2\pi\im}\biggl(\k_1\sum_i\codeg{B_{\text{eff},i}}{p_i+1}\w\dd\deg{\a_i}{p_i}-\frac{\k_2}2\sum_{i,j}\codeg{C_{ij}}{p_i+p_j+2}\w\dd\deg{\a_i}{p_i}\w\dd\deg{\a_j}{p_j}\biggr).
\label{IDalt2}
\end{equation}
This corresponds to an inflow of the anomaly from the bulk CS action, given by
\begin{equation}
\begin{split}
\deg{\I}{D+1}+\deg{\I_\text{c.t.}}{D+1}=\frac1{2\pi\im}\biggl[&\k_1\sum_i\dd\codeg{B_i}{p_i+1}\w\deg{A_i}{p_i+1}\\
&-\frac{\k_2}2\dd\bigl(\codeg{C_{ij}}{p_i+p_j+2}\w\deg{A_i}{p_i+1}\w\deg{A_j}{p_j+1}\bigr)\biggr],
\end{split}
\label{I5}
\end{equation}
which once again leads to the same anomaly polynomial~\eqref{ID2Abelian}. From the variation of the action~\eqref{IDalt2} one can then extract the modified Ward identities for the currents. This time, the conservation laws for $\deg{J_{Ai}}{p_i+1}$ and $\codeg{J_{Bi}}{p_i+1}$ become
\begin{equation}
\dd\ho\deg{J_{Ai}}{p_i+1}\to(-1)^{(D+1)(p_i+1)}\k_1\dd\codeg{B_{\text{eff},i}}{p_i+1},\qquad
\dd\ho\codeg{J_{Bi}}{p_i+1}\to0.
\end{equation}
The conservation law for $\codeg{J_{Cij}}{p_i+p_j+2}$ still keeps the functional form given in eq.~\eqref{Ward_Abelian}, where $\codeg{J_{Bi}}{p_i+1}$ is to be interpreted as the new current, 
$\ho\codeg{J_{Bi}}{p_i+1}\to\k_1\dd\deg{\p_i}{p_i}$.

Alternatively, one may think of $\codeg{B_{\text{eff},i}}{p_i+1}$ in eq.~\eqref{Beff} as a new background gauge field, replacing $\codeg{B_{i}}{p_i+1}$. After all, this would have been the natural choice, had I started from the beginning with the ungauged topological currents, proportional to $\ho\dd\deg{\p_i}{p_i}$ and $\ho(\dd\deg{\p_i}{p_i}\w\dd\deg{\p_j}{p_j})$. The transformation rule for this new source follows from eqs.~\eqref{Beff} and~\eqref{gauge_Abelian},
\begin{equation}
\codeg{B_{\text{eff},i}}{p_i+1}\to\codeg{B_{\text{eff},i}}{p_i+1}+\dd\codeg{\b_{\text{eff},i}}{p_i+2}-\frac{\k_2}{\k_1}\sum_j\codeg{C_{ji}}{p_i+p_j+2}\w\dd\deg{\a_j}{p_j},
\label{Btransfo}
\end{equation}
where
\begin{equation}
\codeg{\b_{\text{eff},i}}{p_i+2}\equiv\codeg{\b_{i}}{p_i+2}-\frac{\k_2}{\k_1}\sum_j\codeg{\g_{ji}}{p_i+p_j+3}\w(\deg{A_j}{p_j+1}+\dd\deg{\a_j}{p_j}).
\end{equation}
In terms of the background field $\codeg{B_{\text{eff},i}}{p_i+1}$, the field strength $\codeg{G_{i}}{p_i}$, defined by eq.~\eqref{FGH_Abelian}, reads
\begin{equation}
\codeg{G_{i}}{p_i}=\dd\codeg{B_{\text{eff},i}}{p_i+1}+\frac{\k_2}{\k_1}\sum_j\dd\codeg{C_{ji}}{p_i+p_j+2}\w\deg{A_j}{p_j+1}.
\end{equation}
The variation of the action~\eqref{IDalt2} then finally gives the new Ward identities for all the currents, where the current $\codeg{J_{Bi}}{p_i+1}$ is now defined by the variation of the action with respect to $\codeg{B_{\text{eff},i}}{p_i+1}$. One finds that by construction, both $\codeg{J_{Bi}}{p_i+1}$ and $\codeg{J_{Cij}}{p_i+p_j+2}$ are conserved, yet they are not gauge-invariant. On the other hand, the $\deg{J_{Ai}}{p_i+1}$ current now becomes manifestly gauge-invariant due to the absence of $\deg{A_j}{p_j+1}$ in the source terms in the action~\eqref{effaction}. It is however not conserved, but rather satisfies the fusion-type Ward identity,
\begin{equation}
\dd\ho\deg{J_{Ai}}{p_i+1}\to(-1)^{(D+1)(p_i+1)}\biggl(\k_1\dd\codeg{B_{\text{eff},i}}{p_i+1}+\frac{\k_2}{\k_1}\dd\codeg{C_{ji}}{p_i+p_j+2}\w\ho\codeg{J_{Bj}}{p_j+1}\biggr).
\end{equation}

This concludes the discussion of alternative presentations of the symmetries of the action and the Ward identities for the currents. As the analysis shows, there are really two different sources of violation of naive current conservation: mixed 't Hooft anomalies and mixing of currents in presence of background gauge fields, arising from the higher-group symmetry. These two effects are independent of each other. On the one hand, ordinary single-component superfluids feature only a 0-form $\gr{U(1)}$ Noether symmetry and its $(D-2)$-form dual together with their mixed anomaly. On the other hand, it will become clear in section~\ref{sec:nonabelian} that it is very well possible to have the same higher-group structure with a strictly invariant action, that is without any 't Hooft anomaly.


\subsection{Some examples}
\label{subsec:examples}

The mechanism whereby a higher-group symmetry arises from the existence of composite currents seems very general, but what are the concrete physical systems where it is actually realized? Thinking for simplicity of a two-component system with NG fields $\deg{\p_1}p$ and $\deg{\p_2}q$, the exterior product $\dd\deg{\p_1}p\w\dd\deg{\p_2}q$ will be the Hodge dual of a new conserved current only if $p+q\leq D-3$. Hence, there is only one possibility in $D=3$ spacetime dimensions, namely $(p,q)=(0,0)$. In $D=4$ spacetime dimensions, the possible choices modulo permutation are $(0,0)$ and $(0,1)$. While one might of course speculate about the physical relevance of more complicated examples in higher dimensions, I will focus on these two cases. 


\subsubsection{\boldmath The $(p,q)=(0,0)$ case}

Choosing the 0-form compact Abelian $\gr{U(1)}$ symmetry, the corresponding Abelian NG field describes an $s$-wave superfluid. Hence, a field theory with two such NG fields naturally represents a two-component superfluid mixture.

Each of the superfluid components possesses a 0-form Noether $\gr{U(1)}$ symmetry corresponding to conservation of particle number, which is spontaneously broken. In the absence of the background gauge fields $\deg{A_i}1$, the associated 1-form Noether current equals, up to normalization, $\dd\p_i$ plus higher-order contributions, depending on the precise equation of state of the system~\cite{Son2002a}. In addition, each superfluid component possesses a dual $(D-2)$-form $\gr{U(1)}$ symmetry with the current $\codeg{J_{Bi}}1$ proportional to $\ho\dd\p_i$, without any higher-order corrections regardless of the interactions. This is closely related to the ``superfluid velocity'' of the $i$-th component, which in the nonrelativistic limit takes the form
\begin{equation}
\vek v_i=\frac{\vek\nabla\p_i}{m_i},
\label{superflow}
\end{equation}
where $m_i$ is the mass of the constituent (atom or molecule) of the $i$-th component. From the general definition~\eqref{charge} it follows that the integral charge associated with the current $\codeg{J_{Bi}}1$ is obtained by integrating $\dd\p_i$ along a closed curve. The dual $(D-2)$-form symmetry therefore measures the winding of the superfluid phase. The corresponding charged object is the worldsheet of a superfluid vortex, which indeed has dimension $D-2$ (the vortex is a point object in $D=3$ and a line object in $D=4$) and gives a nonzero winding number when linked with the closed curve $\S_1$ on which the charge~\eqref{charge} is defined. This $(D-2)$-form symmetry remains unbroken due to the fact that in superfluids, vortices take a finite amount of energy to excite.

Let us now briefly discuss the modification brought about by adding the background gauge fields $\deg{A_i}1$. In the context of superfluids, it is natural to choose the presentation of symmetries worked out in section~\ref{sec:abelian}. Here the Noether $\gr{U(1)}$ symmetries are exact in the sense that the corresponding currents $\deg{J_{Ai}}1$ are exactly conserved. This is expected in an experimental setup where the particle number can be controlled as long as a finite, closed sample of the superfluid is used. Then the dual topological $(D-2)$-form symmetries are necessarily anomalous. The conservation law
\begin{equation}
\dd\ho\codeg{J_{Bi}}1=-\k_1\dd\deg{A_i}1,
\label{dJB}
\end{equation}
being a special case of eq.~\eqref{Ward_Abelian}, has a simple interpretation in charged superfluids, where eq.~\eqref{dJB} tells us that the superfluid winding number is not conserved in the presence of a background magnetic field. This finds a natural realization in the physics of type-II superconductors, where magnetic field can only exist in the presence of vortices. One should however treat this analogy with some care since the basic phenomenology of superconductivity, including the Meissner effect and the very existence of the Abrikosov vortex lattice in type-II superconductors, relies essentially on the dynamical nature of the magnetic field. Here, on the contrary, the $\gr{U(1)}$ gauge fields $\deg{A_i}1$ are treated as a fixed background.

In a two-component superfluid, we now have an additional $(D-3)$-form topological symmetry with current proportional to $\ho(\dd\p_1\w\dd\p_2)$, where the background gauge fields have been set to zero again. In terms of classical field theory, the relation~\eqref{superflow} between the gradient of the NG field and the superflow velocity indicates that the composite current measures the degree to which the superflows of the two superfluid components are noncollinear. The integral charge of the $(D-3)$-form composite symmetry is obtained by integrating $\dd\p_1\w\dd\p_2$ over a two-dimensional surface $\S_2$. Since the group manifold $\gr{U(1)}\times\gr{U(1)}$ has the topology of a torus on which $\dd\p_1\w\dd\p_2$ defines a volume form up to normalization, it is natural to choose $\S_2$ as a torus as well. In $D=4$ spacetime dimensions, the charged object of the composite symmetry can then be taken as the worldsheet of a Hopf link, built out of two vortex rings, one of the $\p_1$ type and the other of the $\p_2$ type. The torus $\S_2$ must be chosen so that it links both vortex rings. The same construction can in principle be repeated in $D=3$ spacetime dimensions, but the interpretation of the resulting charged object is then less clear since it is necessarily localized in time and moreover the two linked rings forming the Hopf link do not represent static vortex rings.


\subsubsection{\boldmath The $(p,q)=(0,1)$ case}

In a QCD-like theory with two light quark flavors of electric charges $q_u,q_d$, each of which carries baryon number $b$, the low-energy dynamics of neutral pions $\pi^0$ (ignoring all other low-energy excitations possibly present) is governed by the following effective Lagrangian in flat four-dimensional Minkowski spacetime~\cite{Brauner2019b},
\begin{equation}
\begin{split}
\La={}&\frac12(\de_\mu\pi^0)^2+m_\pi^2f_\pi^2\cos\biggl(\frac{\pi^0}{f_\pi}\biggr)\\
&-\frac{d}{16\pi^2f_\pi}\epsilon^{\mu\nu\alpha\beta}\bigl[(q_u+q_d)A^Q_{\mu}+2bA^B_{\mu}\bigr]\de_\nu\pi^0\bigl[(q_u-q_d)F^Q_{\alpha\beta}+F^I_{\alpha\beta}\bigr],
\end{split}
\label{QCD}
\end{equation}
where $d$ is the dimension of the representation of the color gauge group that a single quark flavor transforms in. Furthermore, $A_\mu^Q$, $A_\mu^B$ and $A_\mu^I$ are background gauge fields that couple respectively to electric charge, baryon number and isospin, and $F_{\mu\nu}^Q$, $F_{\mu\nu}^B$ and $F_{\mu\nu}^I$ are the corresponding field-strength tensors. Finally, $m_\pi$ and $f_\pi$ are the two effective couplings of the low-energy EFT, representing the pion mass and decay constant.

For physical values of the group charges, $d=3$, $b=1/3$ and $q_u=2/3$, $q_d=-1/3$, the Lagrangian~\eqref{QCD} reproduces the well-known anomalous coupling of the neutral pion to a pair of electromagnetic fields. When the electromagnetic field becomes dynamical, whereas $A_\mu^B$ and $A_\mu^I$ are absent, this is a particular realization of axion electrodynamics~\cite{Wilczek1987a}. The anomalous, topological coupling between the axion (neutral pion) and the electromagnetic field has been shown to give rise to a semistrict 3-group symmetry~\cite{Hidaka2020d,Hidaka2020a,Brennan2020a} that is distinct from the higher-group structure proposed in this paper.

One can, however, choose the group charges differently. Demanding that $q_u+q_d=0$ and dropping for simplicity the background isospin field, the topological coupling of neutral pions to a pair of photons disappears and the Lagrangian~\eqref{QCD} reduces to
\begin{equation}
\La=\frac12(\de_\mu\pi^0)^2+m_\pi^2f_\pi^2\cos\biggl(\frac{\pi^0}{f_\pi}\biggr)-\frac{bd(q_u-q_d)}{8\pi^2f_\pi}\epsilon^{\mu\nu\alpha\beta}A^B_{\mu}\de_\nu\pi^0F^Q_{\alpha\beta}.
\label{QCD2}
\end{equation}
Upon taking the ``chiral limit,'' $m_\pi\to0$, to recover an exact shift symmetry for the neutral pion, and adding a kinetic term for the electromagnetic field, the action stemming from the Lagrangian~\eqref{QCD2} takes precisely the form of eq.~\eqref{masteraction_Abelian} with the $p=0$ NG field corresponding to the neutral pion and the $p=1$ NG field to the photon. The last term in eq.~\eqref{QCD2} corresponds to the source term of the form $\deg C1\w\dd\deg{\p_1}0\w\dd\deg{\p_2}1$. Thus, in the context of QCD, the second-order composite symmetry coupled to the background gauge field $\deg C1$ measures the anomalous baryon number of neutral pions. This is responsible for the appearance of a new phase of QCD in strong magnetic fields, carrying a topological crystalline condensate of neutral pions, dubbed ``chiral soliton lattice''~\cite{Brauner2017a}.

Within a QCD-like theory described by the low-energy effective Lagrangian~\eqref{QCD2}, it is also straightforward to understand what the quantized charged object corresponding to the 0-form composite symmetry coupled to the background field $A_\m^B$ might be. First of all, the continuous shift symmetry acting on $\pi^0$ is explicitly broken by the mass term down to a discrete subgroup. The dual 2-form topological symmetry with current proportional to $\ho\dd\pi^0$ survives. However, its natural charged object is no longer a vortex, but rather a domain wall, interpolating between the neighboring minima of the potential in eq.~\eqref{QCD2}. Going somewhat beyond the discussion in section~\ref{subsec:charges}, the topological charge of the domain wall is obtained by integrating $\dd\pi^0$ over an \emph{open} curve $\S_1$ piercing the domain wall. The charged object of the magnetic 1-form symmetry with current proportional to $\ho F^Q$ is the worldline of a magnetic monopole, as explained in section~\ref{subsec:objects}, and its topological charge is measured by integrating $F^Q$ over a closed surface $\S_2$ surrounding the monopole. The charged object of the composite symmetry is then constructed out of a monopole and a domain wall that are topologically linked. The configuration that does the job is one where the monopole is surrounded by a spherical domain wall; this kind of soliton was considered previously in relation to the so-called Witten effect~\cite{Hidaka2020d}. The charge of the composite object is obtained by integrating $\dd\pi^0\w F^Q$ over the three-dimensional domain $\S_1\times\S_2$.

It is interesting to compare the $(p,q)=(0,1)$ case of our general setup to the Goldstone-Maxwell (GM) model, shown in ref.~\cite{Cordova2019a} to possess a global 3-group symmetry. In order to make the notations match, one needs to make the following replacements, where the right-hand side indicates the notation used by C\'ordova et al.,
\begin{equation}
\begin{gathered}
\deg{\p_1}0\to\chi,\qquad
\deg{A_1}1\to\deg A1,\qquad
\deg{B_{\text{eff},1}}3\to\deg\Theta3,\\
\deg{\p_2}1\to\deg c1,\qquad
\deg{A_2}2\to\deg{B_\text{e}}2,\qquad
\deg{B_{\text{eff},2}}2\to\deg{B_\text{m}}2,\\
\deg{\a_1}0\to\deg{\l_A}0,\qquad
\deg{\a_2}1\to\deg{\Lambda_\text{e}}1,\qquad
\deg{\b_{\text{eff},1}}2\to\deg{\Lambda_\Theta}2,\qquad
\deg{\b_{\text{eff},2}}1\to\deg{\Lambda_\text{m}}1.
\end{gathered}
\end{equation}
Then the action~\eqref{effaction} recovers that of the GM model \emph{provided} one simultaneously identifies the background field $\deg{C_{12}}1\equiv\deg C1$ with $\deg A1$ and chooses the couplings as
\begin{equation}
\k_1=\frac\im{2\pi},\qquad
\k_2=-\frac{\im\hat\kappa_A}{4\pi^2},
\end{equation}
where $\hat\kappa_A$ is the 2-group coupling of the GM model. The transformation rules~\eqref{Btransfo} agree with those displayed in ref.~\cite{Cordova2019a} up to a total derivative term absorbed into a redefinition of $\codeg{\b_{\text{eff},i}}{p_i+2}$, and the 5-form~\eqref{I5} reproduces the bulk CS action shown in eq.~(6.69) of ref.~\cite{Cordova2019a}.

This comparison reveals a possible mechanism to generate further theories with higher-group symmetry beyond those introduced in this paper. Namely, it is in principle possible to ``contract'' the underlying higher-group symmetry by identifying some of the background fields (and the corresponding transformation parameters), as long as this is consistent with all the transformation rules. It would be interesting to investigate if one can obtain more physically relevant field theories in this way.


\section{Generalization to higher-order composite currents}
\label{sec:abelianhigher}

In this section, I will generalize the construction put forward above by taking into account the possibility that a given system possesses composite topological currents of higher order than two. I should stress right at the outset that the motivation for doing so is theoretical rather than practical. Namely, there cannot be any nontrivial composite topological symmetries of order $D$ or higher in $D$ spacetime dimensions. The first, minimal example of a higher-order composite symmetry therefore appears in $D=4$ and can be thought of as a three-component superfluid where all the primitive symmetries are 0-form.

On the other hand, the extension of the formalism developed so far reveals a rather elegant underlying structure, where the deformation of the transformation rule for $\codeg{B_i}{p_i+1}$ displayed in eq.~\eqref{gauge_Abelian} repeats iteratively at higher orders. Due to the above-mentioned practical limitations, I will however restrict the generality of the discussion by the assumption that all the (still Abelian) Noether symmetries present in the system are 0-form. This makes it possible to avoid having to deal with complicated mixed symmetries of the higher-order topological currents under permutation of their constituents.

The basic ingredients in the setup are thus a set of 0-form NG fields $\p_i$ subject to independent shift symmetries, along with the associated 1-form background gauge fields $\deg{A_i}1$. These will only enter the action through the combinations $\deg{\Om_i}1=\dd\p_i-\deg{A_i}1$, invariant under the simultaneous gauge transformation~\eqref{Noethersym}, where $p_i=0$ and accordingly $\a_i$ is a 0-form just like $\p_i$ itself. Out of these covariant derivatives, one can construct $n$-th order composite topological currents as $\ho\bigl(\deg{\Om_{i_1}}1\w\dotsb\w\deg{\Om_{i_n}}1\bigr)$ up to an arbitrary normalization factor. These currents can in turn be coupled to a set of background gauge fields $\codeg{B_{i_1\dotsb i_n}}n$. Both the currents and the gauge fields are fully antisymmetric in their flavor indices.

A generic action featuring all the NG fields and background fields then takes a form analogous to eq.~\eqref{masteraction_Abelian},
\begin{equation}
S=S_\text{inv}+\int\sum_{n=1}^{D-1}\frac{\k_n}{n!}\codeg{B_{i_1\dotsb i_n}}{n}\w(\dd\p_{i_1}-\deg{A_{i_1}}1)\w\dotsb\w(\dd\p_{i_n}-\deg{A_{i_n}}1).
\label{masteraction_higherorder}
\end{equation}
Now that it is not necessary to label all the constituents with a generic degree, it is safe to use Einstein's summation convention. The sum over source terms for the composite currents runs in principle up to the maximum possible degree of $D-1$, but may terminate earlier depending on the number of independent primitive currents in the theory. A naive Abelian transformation of the background fields $\codeg{B_{i_1\dotsb i_n}}n$ leads to an undesired, $\p_i$-dependent shift of the action at every order of the sum. This issue can, however, be fixed by using the following set of simultaneous gauge transformations,
\begin{equation}
\begin{split}
\p_i&\to\p_i+\a_i,\\
\deg{A_i}1&\to\deg{A_i}1+\dd\a_i,\\
\codeg{B_{i_1\dotsb i_n}}n&\to\codeg{B_{i_1\dotsb i_n}}n+\dd\codeg{\b_{i_1\dotsb i_n}}{n+1}+(-1)^{D+1}\frac{\k_{n+1}}{\k_n}\codeg{\b_{i_1\dotsb i_n i_{n+1}}}{n+2}\w\dd\deg{A_{i_{n+1}}}1,
\end{split}
\label{gauge_higherorder}
\end{equation}
which is an obvious generalization of eq.~\eqref{gauge_Abelian}.\footnote{The last background field in the hierarchy transforms merely as $\codeg{B_{i_1\dotsb i_{D-1}}}{D-1}\to\codeg{B_{i_1\dotsb i_{D-1}}}{D-1}+\dd\codeg{\b_{i_1\dotsb i_{D-1}}}{D}$, but this can be included in eq.~\eqref{gauge_higherorder} by formally setting $\k_D=0$ therein. An analogous remark applies to eqs.~\eqref{FGHhigher} and~\eqref{FGHhigherorder} below: $\k_D$ is implicitly understood to be zero wherever it appears.} Upon this gauge transformation, the action shifts by
\begin{equation}
\d S=\int(-1)^{D}\k_1\codeg{\b_i}{2}\w\dd\deg{A_i}{1},
\label{dS_higherorder}
\end{equation}
where I have again discarded surface terms generated by the transformation, see footnote~\ref{ftn:surface} on page~\pageref{ftn:surface}. It is amusing to point out that this anomalous variation of the action can be eliminated altogether~\cite{Cordova2020b} if one formally extends the sum over source terms in eq.~\eqref{masteraction_higherorder} to $n=0$. This amounts to adding a $D$-form background field $\codeg{B}0$ to the action through a tadpole term $\k_0\codeg{B}0$, and to equipping the new field with the transformation rule $\codeg{B}0\to\codeg{B}0+\dd\codeg{\b}1+(-1)^{D+1}\frac{\k_1}{\k_0}\codeg{\b_i}2\w\dd\deg{A_i}1$, which likewise extends the transformation rule~\eqref{gauge_higherorder} to the $n=0$ case. The 't Hooft anomaly indicated by eq.~\eqref{dS_higherorder} of course does not disappear, it is merely swept under the rug, or more accurately moved from the right-hand side of eq.~\eqref{dS_higherorder} to the left-hand side.

The deformation of the transformation rules for the background gauge fields as compared to the naively expected Abelian shifts requires an appropriate modification of the corresponding field strengths. It is easy to check that the following definitions do the job, generalizing eq.~\eqref{FGH_Abelian} to the whole hierarchy of higher-order topological symmetries,
\begin{equation}
\begin{split}
\deg{F_i}2&=\dd\deg{A_i}1,\\
\codeg{G_{i_1\dotsb i_n}}{n-1}&=\dd\codeg{B_{i_1\dotsb i_n}}n+(-1)^{D}\frac{\k_{n+1}}{\k_n}\codeg{B_{i_1\dotsb i_n i_{n+1}}}{n+1}\w\dd\deg{A_{i_{n+1}}}1.
\end{split}
\label{FGHhigher}
\end{equation}
The leads in turn to modified Bianchi identities,
\begin{equation}
\begin{split}
\dd\deg{F_i}2&=0,\\
\dd\codeg{G_{i_1\dotsb i_n}}{n-1}&=(-1)^D\frac{\k_{n+1}}{\k_n}\codeg{G_{i_1\dotsb i_ni_{n+1}}}{n}\w\deg{F_{i_{n+1}}}2.
\end{split}
\label{FGHhigherorder}
\end{equation}
With the gauge transformation rules~\eqref{gauge_higherorder} and the variation of the action~\eqref{dS_higherorder} at hand, it is straightforward to extract the Ward identities for all the conserved currents of the theory. With the definition of the currents,
\begin{equation}
\ho\deg{J_{Ai}}1\equiv\frac{\d S}{\d\deg{A_i}1},\qquad
\ho\codeg{J_{Bi_1\dotsb i_n}}{n}\equiv\frac{\d S}{\d\codeg{B_{i_1\dotsb i_n}}n},
\end{equation}
one finds upon a few lines of manipulation the following,
\begin{align}
\notag
\dd\ho\deg{J_{Ai}}1&=0,\\
\dd\ho\codeg{J_{Bi}}1&=-\k_1\dd\deg{A_i}1,\\
\notag
\dd\ho\codeg{J_{Bi_1\dotsb i_n}}n&=(-1)^n\frac{\k_n}{\k_{n-1}}\sum_{\text{cyclic }\pi}\sgn\pi\,\dd\deg{A_{i_{\pi(n)}}}1\w\ho\codeg{J_{Bi_{\pi(1)}\dotsb i_{\pi(n-1)}}}{n-1}\qquad\text{for $2\leq n\leq D-1$},
\end{align}
where on the last line the sum runs over all cyclic permutations $\pi$ of the flavor indices.

The discussion of possible alternative presentations of the symmetry in this general setup including the whole hierarchy of higher-order topological currents would run in a close analogy to section~\ref{subsec:ambiguities}, and I will not go to details. Let me just stress that the action~\eqref{masteraction_higherorder} was \emph{chosen} so that the Noether currents $\deg{J_{Ai}}1$ are by construction conserved, although not gauge-invariant. The currents $\codeg{J_{Bi_1\dotsb i_n}}n$, on the other hand, are by construction gauge-invariant, yet not conserved. The anomalous variation of the action~\eqref{dS_higherorder} can be understood as arising from anomaly inflow due to a CS theory, living in the $(D+1)$-dimensional bulk,
\begin{equation}
\deg\I{D+1}=\frac{\k_1}{2\pi\im}(-1)^{D}\codeg{B_i}{1}\w\dd\deg{A_i}{1}.
\label{ID1higherorder}
\end{equation}
Taking an exterior derivative then leads in turn to a gauge-invariant $(D+2)$-form anomaly polynomial,
\begin{equation}
\deg\I{D+2}=\frac{\k_1}{2\pi\im}(-1)^{D}\dd\codeg{B_i}{1}\w\dd\deg{A_i}{1}=\frac{\k_1}{2\pi\im}(-1)^{D}\codeg{G_i}{0}\w\deg{F_i}{2},
\label{ID2higherorder}
\end{equation}
where the definition~\eqref{FGHhigherorder} of the field strength $D$-form $\codeg{G_i}0$ has been used.


\section{Dual description of the higher-group symmetry}
\label{sec:duality}

Abelian superfluids have a dual description in which the $p$-form NG field $\deg{\p}{p}$ is interchanged with its Hodge dual, and likewise the background fields $\deg{A}{p+1}$ and $\codeg{B}{p+1}$ are swapped (see ref.~\cite{Hjelmeland1997a} for an introduction). This duality is important for understanding the dynamics of vortices in a superfluid medium. In $D=3$ spacetime dimensions, it actually represents just one link in a larger network of dualities between various field theories, see for instance ref.~\cite{Karch2016a} and references therein.

Here I have the modest goal of using the duality to offer a different perspective on the higher-group symmetry of field theories with composite currents. This will also implicitly underline the fact that the same physical system can have quite different presentations based on different local degrees of freedom, possibly featuring different gauge redundancies. What has to remain the same in such different descriptions are the global symmetries, encoded in the background gauge invariance of the generating functional, including their possible 't Hooft anomalies~\cite{Gaiotto2015a}.

In order to achieve this goal, I will once again restrict myself to the simplest situation where only second-order composite currents are taken into account. The starting point of the duality transformation of the action~\eqref{masteraction_Abelian} is the observation that the NG fields $\deg{\p_i}{p_i}$ only enter it through their exterior derivatives, $\dd\deg{\p_i}{p_i}$. One may wish to cast the theory in terms of these ``currents,'' closely related but not identical to the topological currents $\codeg{J_{Bi}}{p_i+1}$. This is done by introducing a ``parent action'' with dynamical variables $\deg{\J_i}{p_i+1}$ and $\codeg{\c_i}{p_i+2}$,
\begin{equation}
\begin{split}
S=S_\text{inv}+\int\biggl\{&\k_1\sum_i\Bigl[\codeg{B_i}{p_i+1}\w(\deg{\J_i}{p_i+1}-\deg{A_i}{p_i+1})-\dd\codeg{\c_i}{p_i+2}\w\deg{\J_i}{p_i+1}\Bigr]\\
&+\frac{\k_2}2\sum_{i,j}\codeg{C_{ij}}{p_i+p_j+2}\w(\deg{\J_i}{p_i+1}-\deg{A_i}{p_i+1})\w(\deg{\J_j}{p_j+1}-\deg{A_j}{p_j+1})\biggr\},
\end{split}
\label{parentaction}
\end{equation}
where $\dd\deg{\p_i}{p_i}$ has been replaced with $\deg{\J_i}{p_i+1}$ also in the invariant part of the action, $S_\text{inv}$. The new ``vortex variables'' $\codeg{\c_i}{p_i+2}$ act as Lagrange multipliers for the constraint $\dd\deg{\J_i}{p_i+1}=0$. Indeed, upon integrating $\codeg{\c_i}{p_i+2}$ out by using their equation of motion, the ``current variables'' $\deg{\J_i}{p_i+1}$ are forced to become closed, and once they are cast as $\deg{\J_i}{p_i+1}=\dd\deg{\p_i}{p_i}$ by virtue of the Poincar\'e lemma, the original action~\eqref{masteraction_Abelian} is recovered. The point of this exercise is that one may wish to integrate out $\deg{\J_i}{p_i+1}$ instead of $\codeg{\c_i}{p_i+2}$. By doing so, one arrives at an EFT formulated in terms of the vortex variables, coupled to the same set of background fields as the NG fields $\deg{\p_i}{p_i}$ in the original theory~\eqref{masteraction_Abelian}. In fact, one can choose for each flavor $i$ separately to integrate out either the current or the vortex variable. In a system with $n$ different flavors of NG bosons, we then have in principle $2^n$ different representations of the same system.

In order to get insight into how the symmetry of the system is represented in all these different mutations, it is sufficient to understand how it is realized on the parent action~\eqref{parentaction}. The transformation rules for the background fields must of course remain the same as in eq.~\eqref{gauge_Abelian}; the background symmetry of any given theory cannot depend on the choice of its dynamical variables. What we however need is the transformation rules for $\deg{\J_i}{p_i+1}$ and $\codeg{\c_i}{p_i+2}$. The former can be guessed to be a simple shift by $\dd\deg{\a_i}{p_i}$ thanks to the fact that the parent action~\eqref{parentaction} only depends on the difference $\deg{\J_i}{p_i+1}-\deg{A_i}{p_i+1}$. The latter can be guessed by trial and error. It turns out that the variation of the action~\eqref{dS_Abelian} is correctly reproduced, modulo surface terms, if one uses the following prescription,
\begin{equation}
\begin{split}
\deg{\J_i}{p_i+1}&\to\deg{\J_i}{p_i+1}+\dd\deg{\a_i}{p_i},\\
\codeg{\c_i}{p_i+2}&\to\codeg{\c_i}{p_i+2}+\codeg{\b_i}{p_i+2}+\frac{\k_2}{\k_1}\sum_j\codeg{\g_{ji}}{p_i+p_j+3}\w(\deg{\J_j}{p_j+1}-\deg{A_j}{p_j+1}).
\end{split}
\label{gaugeJchi}
\end{equation}

Three comments are in order here. First, in the absence of composite currents, or of the background field $\codeg{C_{ij}}{p_i+p_j+2}$, the particle-vortex duality is perfect: $\deg{\p_i}{p_i}$ and $\deg{A_i}{p_i+1}$ transform by a shift with respect to $\deg{\a_i}{p_i}$, whereas $\codeg{\c_i}{p_i+2}$ and $\codeg{B_i}{p_i+1}$ transform by a shift with respect to $\codeg{\b_i}{p_i+2}$. The only element that breaks the symmetry between the two descriptions of the superfluid is the mixed 't Hooft anomaly of the Noether symmetries and their duals. This decides whether it is the particle current $\deg{J_{Ai}}{p_i+1}$ or the vortex current $\codeg{J_{Bi}}{p_i+1}$ that is conserved. Coupling in the composite currents via adding the background fields $\codeg{C_{ij}}{p_i+p_j+2}$ makes the action of the symmetry on the vortex variables highly nontrivial, and the above perfect duality is lost.

Second, the realization of the symmetry in terms of the $\deg{\J_i}{p_i+1}$ and $\codeg{\c_i}{p_i+2}$ variables is independent of the choice of the invariant part of the action, $S_\text{inv}$, and given by eq.~\eqref{gaugeJchi}. This is however no longer true when the current variables are integrated out and the transformation of $\codeg{\c_i}{p_i+2}$ is to be expressed solely in terms of the $\codeg{\c_j}{p_j+2}$s and the background gauge fields. The resulting action of the symmetry on $\codeg{\c_i}{p_i+2}$ will depend on the choice of $S_\text{inv}$ very sensitively and cannot in general be given in a closed form.

Third, the duality transformation helps to shed some light on higher-form symmetries, should one wish to think of them in terms of transformations of local degrees of freedom. Indeed, as eq.~\eqref{gauge_Abelian} makes clear, the $(D-p_i-2)$-form symmetry with parameter $\codeg{\b_i}{p_i+2}$ only acts on the background fields. The Noether symmetry with parameter $\deg{\a_i}{p_i}$ may also be higher-form in case $p_i\geq1$, but obviously needs the NG field $\deg{\p_i}{p_i}$ to be realized as a local transformation. Introducing the dual variable $\codeg{\c_i}{p_i+2}$ makes it possible to realize the dual $(D-p_i-2)$-form symmetry locally as well. The trick of course is that the relationship between $\deg{\p_i}{p_i}$ and $\codeg{\c_i}{p_i+2}$ itself is nonlocal.

Before I show a concrete illustration of the action of the symmetry on the EFT formulated in terms of the vortex variables, let me briefly mention that the above expressions for the parent action~\eqref{parentaction} and the symmetry transformation~\eqref{gaugeJchi} can be extended to the setup of section~\ref{sec:abelianhigher} where the whole hierarchy of higher-order composite currents is taken into account, albeit with the simplifying assumption that all the Noether symmetries, and thus also the $\p_i$ variables, are 0-forms. In this case, the parent action can be written as
\begin{equation}
S=S_\text{inv}+\int\biggl[-\k_1\dd\codeg{\c_i}{2}\w\deg{\J_i}1+\sum_{n=1}^{D-1}\frac{\k_n}{n!}\codeg{B_{i_1\dotsb i_n}}{n}\w(\deg{\J_{i_1}}1-\deg{A_{i_1}}1)\w\dotsb\w(\deg{\J_{i_n}}1-\deg{A_{i_n}}1)\biggr].
\end{equation}
The variation of the action~\eqref{dS_higherorder} is correctly reproduced if the transformation of the background gauge fields as indicated in eq.~\eqref{gauge_higherorder} is augmented with the following transformations of the current variables $\deg{\J_i}1$ and the vortex variables $\codeg{\c_i}2$,
\begin{equation}
\begin{split}
\deg{\J_i}1&\to\deg{\J_i}1+\dd\a_i,\\
\codeg{\c_i}2&\to\codeg{\c_i}2+\codeg{\b_i}2+\frac1{\k_1}\sum_{n=1}^{D-2}\frac{\k_{n+1}}{n!}\codeg{\b_{j_1\dotsb j_n i}}{n+2}\w(\deg{\J_{j_1}}1-\deg{A_{j_1}}1)\w\dotsb\w(\deg{\J_{j_n}}1-\deg{A_{j_n}}1).
\end{split}
\end{equation}
The $\codeg{\b_i}2$ term can actually be included in the sum over $n$ by formally extending the latter to $n=0$. Obviously, adding higher-order composite currents makes the transformation law for the vortex variables even more complicated. In general, if the maximum order of composite current included is $n$, then $\codeg{\c_i}2$ shifts by a polynomial of order $n-1$ in the current variables.


\subsection{Explicit example with two 0-form symmetries}
\label{subsec:dualexample}

Let us now have a look at a concrete example. In order to keep things as simple as possible, I will consider a noninteracting two-component superfluid in a flat three-dimensional Euclidean spacetime. In this case, the parent action~\eqref{parentaction}, now including explicitly the invariant part, reduces to
\begin{equation}
\begin{split}
S=\int\biggl[&\frac12(\deg{\J_i}1-\deg{A_i}1)\w\ho(\deg{\J_i}1-\deg{A_i}1)+\k_1\deg{B_i}2\w(\deg{\J_i}1-\deg{A_i}1)-\k_1\dd\deg{\c_i}1\w\deg{\J_i}1\\&+\k_2\deg C1\w(\deg{\J_1}1-\deg{A_1}1)\w(\deg{\J_2}1-\deg{A_2}1)\biggr],
\end{split}
\label{parent_example}
\end{equation}
where the implicit sum runs over $i=1,2$, and in the last term, I used the antisymmetry of $\deg{C_{ij}}1$ in its flavor indices to replace it with $\epsilon_{ij}\deg C1$.

This theory has altogether four equivalent presentations depending on the independent choice of the particle or vortex variable as the dynamical degree of freedom for each of the superfluid flavors. In case one integrates out both vortex variables $\deg{\c_i}1$, one gets back to the standard description of the superfluid in terms of a shift-invariant action for the NG fields~$\p_i$, see eq.~\eqref{masteraction_Abelian}. 

Let us now consider a hybrid description in terms of, say, $\p_1$ and $\deg{\c_2}1$. To that end, we need to integrate out the vortex variable $\deg{\c_1}1$ and the current variable $\deg{\J_2}1$. Integrating out $\deg{\c_1}1$ leads to the constraint $\dd\deg{\J_1}1=0$ and in turn to the representation $\deg{\J_1}1=\dd\p_1$. The equation of motion for $\deg{\J_2}1$, on the other hand, reads
\begin{equation}
\deg{\J_2}1-\deg{A_2}1=\k_1\ho(\dd\deg{\c_2}1-\deg{B_2}2)-\k_2\ho[\deg C1\w(\deg{\J_1}1-\deg{A_1}1)].
\label{EoMJ1}
\end{equation}
This just needs to be plugged back into the action~\eqref{parent_example}. Upon a short manipulation, one arrives at the final expression for the action,
\begin{equation}
\begin{split}
S=\int\biggl[&\frac12\bigl|\dd\p_1-\deg{A_1}1\bigr|^2-\frac12\bigl|\k_1(\dd\deg{\c_2}1-\deg{B_2}2)-\k_2\deg C1\w(\dd\p_1-\deg{A_1}1)\bigr|^2\\
&+\k_1\deg{B_1}2\w(\dd\p_1-\deg{A_1}1)-\k_1\dd\deg{\c_2}1\w\deg{A_2}1\biggr],
\end{split}
\end{equation}
with the shorthand notation $|\deg\om p|^2\equiv\deg\om p\w\ho\deg\om p$. The representation of the symmetry on the dynamical fields in this action descends from that on the parent action, eq.~\eqref{gaugeJchi},
\begin{equation}
\begin{split}
\p_1&\to\p_1+\a_1,\\
\deg{\c_2}1&\to\deg{\c_2}1+\deg{\b_2}1+\frac{\k_2}{\k_1}\g(\dd\p_1-\deg{A_1}1),
\end{split}
\label{parent_example_hybrid}
\end{equation}
where $\g$ is the 0-form parameter of the gauge transformation $\deg C1\to\deg C1+\dd\g$. This shows that integrating out the current variable $\deg{\J_2}1$ makes the dual 1-form symmetry with the parameter $\deg{\b_2}1$ into a nontrivial contact symmetry whereby the shift of $\deg{\c_2}1$ depends on the (covariant) derivative of $\p_1$, or in physical terms the superfluid velocity of the first superfluid component.

The descendant action~\eqref{parent_example_hybrid} also nicely illustrates a general feature, pointed out in the introduction, namely that which current is Noether --- that is requires for its conservation the equations of motion --- and which is conserved identically depends on the choice of dynamical degrees of freedom. In the presentation of the action~\eqref{masteraction_Abelian} in terms of the NG degrees of freedom, the $\deg{J_{Ai}}{p_i+1}$ currents are Noether, whereas $\codeg{J_{Bi}}{p_i+1}$ and $\codeg{J_{Cij}}{p_i+p_j+2}$ are topological. Now the only topological currents are
\begin{equation}
\ho\deg{J_{A2}}1=-\k_1\dd\deg{\c_2}1,\qquad
\ho\deg{J_{B1}}2=\k_1(\dd\p_1-\deg{A_1}1).
\end{equation}
All the remaining three currents, $\deg{J_{A1}}1$, $\deg{J_{B2}}2$ and $\deg{J_C}1$, couple to background fields that appear in the kinetic terms in the action~\eqref{parent_example_hybrid}, and their conservation thus essentially depends on the equations of motion. In case of $\deg{J_C}1$ this is really a secondary effect: the Ward identity for $\deg{J_C}1$ stemming from eq.~\eqref{Ward_Abelian} depends crucially on that for $\deg{J_{B2}}2$, but the Ward identity for $\deg{J_{B2}}2$ requires the equation of motion for $\deg{\c_2}1$.

If keeping one of the vortex variables makes things complicated, one can expect the realization of the symmetry to become even more nontrivial in the fully dualized description where both current variables $\deg{\J_i}1$ are integrated out. Then the equation of motion for $\deg{\J_2}1$ is augmented with that for $\deg{\J_1}1$,
\begin{equation}
\deg{\J_1}1-\deg{A_1}1=\k_1\ho(\dd\deg{\c_1}1-\deg{B_1}2)+\k_2\ho[\deg C1\w(\deg{\J_2}1-\deg{A_2}1)].
\label{EoMJ2}
\end{equation}
Equations~\eqref{EoMJ1} and~\eqref{EoMJ2} together make a set of linear equations for the two current variables $\deg{\J_i}1$. It is here that the assumption of a flat three-dimensional spacetime is needed. The solution can be written jointly for both currents,
\begin{equation}
\begin{split}
\deg{\J_i}1-\deg{A_i}1=\frac{\k_1}{1-\k_2^2C^2}\Bigl\{&\ho(\dd\deg{\c_i}1-\deg{B_i}2)+\k_2\epsilon_{ij}\ho\bigl[\deg C1\w\ho(\dd\deg{\c_j}1-\deg{B_j}2)\bigr]\\
&-\k_2^2\deg C1\w\ho\bigl[\deg C1\w(\dd\deg{\c_i}1-\deg{B_i}2)\bigr]\Bigr\},
\end{split}
\end{equation}
where $C^2$ is a shorthand notation for $\ho(\deg C1\w\ho\deg C1)$. The resulting action for the vortex variables $\deg{\c_i}1$ follows by inserting this in eq.~\eqref{parent_example}, and I will not write it out explicitly. The transformation rule for
$\deg{\c_i}1$ likewise descends from eq.~\eqref{gaugeJchi} and is given implicitly by
\begin{equation}
\deg{\c_i}1\to\deg{\c_i}1+\deg{\b_i}1-\frac{\k_2}{\k_1}\g\epsilon_{ij}(\deg{\J_j}{1}-\deg{A_j}{1}).
\end{equation}
Here we thus end up with a realization of both 1-form symmetries as contact symmetries, whereby the transformation of each of $\deg{\c_i}1$ depends on the exterior derivative of both of $\deg{\c_i}1$. One could have hardly expected to be able to reveal the underlying higher-group structure directly, without starting from the description in terms of the NG fields $\p_i$.


\section{Generalization to non-Abelian symmetries}
\label{sec:nonabelian}

In this section, I will outline a different generalization of the basic setup introduced in section~\ref{sec:abelian}. Namely, as explained in section~\ref{sec:topcurrents}, in systems with spontaneously broken symmetry, additional, emergent topological symmetries naturally arise from the geometry of the coset space on which the NG degrees of freedom live. The Abelian setup of section~\ref{sec:abelian} is a special case, whereby every Abelian NG field $\deg{\p}p$ gives rise to a topological current, equal $\ho\dd\deg\p p$ up to a normalization factor.

I will now allow for the possibility that the Noether symmetry whose spontaneous breakdown is responsible for the existence of the NG modes may be non-Abelian. Owing to the fact that higher-form symmetries are naturally Abelian~\cite{Gaiotto2015a}, I will simplify the notation by only taking into account 0-form Noether symmetries from the outset. The low-energy dynamics is then governed by a set of NG scalars, $\pi^a$, that parametrize the coset space $G/H$ of the spontaneously broken 0-form symmetry. Associated with the whole 0-form symmetry group $G$, there is a set of background gauge fields, conveniently merged into a matrix 1-form connection, $\deg A1\equiv\deg A1{}^iT_i$.

As shown in section~\ref{sec:topcurrents}, a class of emergent topological symmetries is associated with the de Rham cohomology of the coset space $G/H$. Explicit forms for cohomology generators of degree one to three were given in eqs.~\eqref{cohom1}, \eqref{cohom2} and~\eqref{cohom3}. Here I will make the general assumption that the given theory possesses a set of globally well-defined $p_\A$-forms $\deg{\Om_\A}{p_\A}$ with the properties that (i) they are gauge-invariant under the simultaneous gauge transformation of the NG fields and the background gauge field $\deg A1$ of $G$, and (ii) $\dd\deg{\Om_\A}{p_\A}$ is a local function of $\deg A1$ alone, independent of the NG fields.\footnote{In section~\ref{sec:abelian} the same indices $i,j,\dotsc$ were used for the primitive topological currents as in section~\ref{sec:topcurrents} for the generators of the symmetry group $T_{i,j,\dotsc}$. This is consistent if one deals with completely broken Abelian symmetry. In the present section, a different set of indices is needed, and I will use the lowercase fraktur letters $\A,\B,\dotsc$} Both of these properties are satisfied by the (gauged) cohomology generators listed in section~\ref{sec:topcurrents}.

The Hodge duals $\ho\deg{\Om_\A}{p_\A}$ play the role of the topological currents. Associated with each of them, there is a background gauge field $\codeg{B_\A}{p_\A}$. Analogously to section~\ref{sec:abelian}, I will in addition allow for the second-order topological currents $\ho\bigl(\deg{\Om_\A}{p_\A}\w\deg{\Om_\B}{p_\B}\bigr)$, and couple them to a set of background gauge fields $\codeg{C_{\A\B}}{p_\A+p_\B}$ with the symmetry property
\begin{equation}
\codeg{C_{\B\A}}{p_\A+p_\B}=(-1)^{p_\A p_\B}\codeg{C_{\A\B}}{p_\A+p_\B}.
\end{equation}
A generic action including such primitive and second-order composite topological currents then takes the form
\begin{equation}
S=S_\text{inv}+\int\Bigl(\k_1\sum_\A\codeg{B_\A}{p_\A}\w\deg{\Om_\A}{p_\A}+\frac{\k_2}2\sum_{\A,\B}\codeg{C_{\A\B}}{p_\A+p_\B}\w\deg{\Om_\A}{p_\A}\w\deg{\Om_\B}{p_\B}\Bigr).
\label{masteraction_nonAbelian}
\end{equation}

Since the basic strategy for analyzing the background gauge symmetry was settled in the previous sections, I can go through the individual steps rather quickly. First, in order to get rid of undesired, NG-field-dependent terms in the variation of the action, the naively expected transformations of the background fields have to be modified to
\begin{equation}
\begin{split}
\deg A1&\to g\deg A1g^{-1}+g\dd g^{-1}\qquad\text{where $g\in G$},\\
\codeg{B_\A}{p_\A}&\to\codeg{B_\A}{p_\A}+\dd\codeg{\b_\A}{p_\A+1}+\frac{\k_2}{\k_1}\sum_\B(-1)^{D+(p_\A+1)(p_\B+1)}\codeg{\g_{\A\B}}{p_\A+p_\B+1}\w\dd\deg{\Om_\B}{p_\B},\\
\codeg{C_{\A\B}}{p_\A+p_\B}&\to\codeg{C_{\A\B}}{p_\A+p_\B}+\dd\codeg{\g_{\A\B}}{p_\A+p_\B+1},
\end{split}
\label{gauge_nonAbelian}
\end{equation}
where the first line is simply copied from eq.~\eqref{Atransfo}. Note that thanks to the fact that $\dd\deg{\Om_\A}{p_\A}$ only depends on the 1-form background gauge fields of $G$, the transformation of $\codeg{B_\A}{p_\A}$ is well-defined. Under the simultaneous transformations~\eqref{gauge_nonAbelian}, the action~\eqref{masteraction_nonAbelian} changes, up to surface terms, by
\begin{equation}
\d S=\int\sum_\A(-1)^{D+p_\A}\k_1\codeg{\b_\A}{p_\A+1}\w\dd\deg{\Om_\A}{p_\A}.
\label{dS_nonAbelian}
\end{equation}
This is only nonvanishing, leading to a 't Hooft anomaly, for cohomology generators $\deg{\Om_\A}{p_\A}$ of odd degrees, whose full gauging under $G$ faces an obstruction, related to the chiral anomaly. Topological currents based on cohomology generators of even degrees can be added to the action without affecting its full gauge invariance under $G$.

The modified transformation laws~\eqref{gauge_nonAbelian} require an appropriate modification of the field strength for $\codeg{B_\A}{p_\A}$,
\begin{equation}
\begin{split}
\deg F2&=\dd\deg A1+\deg A1\w\deg A1,\\
\codeg{G_\A}{p_\A-1}&=\dd\codeg{B_\A}{p_\A}-\frac{\k_2}{\k_1}\sum_\B(-1)^{D+(p_\A+1)(p_\B+1)}\codeg{C_{\A\B}}{p_\A+p_\B}\w\dd\deg{\Om_\B}{p_\B},\\
\codeg{H_{\A\B}}{p_\A+p_\B-1}&=\dd\codeg{C_{\A\B}}{p_\A+p_\B}.
\end{split}
\label{FGH_nonAbelian}
\end{equation}
This in turn leads to a modified Bianchi identity for $\codeg{G_\A}{p_\A-1}$. For the sake of completeness, I list the Bianchi identities for all the symmetries,
\begin{equation}
\begin{split}
\dd\deg F2&=\deg F2\w\deg A1-\deg A1\w\deg F2,\\
\dd\codeg{G_\A}{p_\A-1}&=-\frac{\k_2}{\k_1}\sum_\B(-1)^{D+(p_\A+1)(p_\B+1)}\codeg{H_{\A\B}}{p_\A+p_\B-1}\w\dd\deg{\Om_\B}{p_\B},\\
\dd\codeg{H_{\A\B}}{p_\A+p_\B-1}&=0.
\end{split}
\label{Bianchi_nonAbelian}
\end{equation}

Once a concrete action is known, explicit expressions for the currents that the background gauge fields couple to can be obtained as usual by taking variation of the action with respect to the background fields,
\begin{equation}
\ho\deg{J_{A,i}}{1}\equiv\frac{\d S}{\d\deg{A}{1}{}^i},\qquad
\ho\codeg{J_{B,\A}}{p_\A}\equiv\frac{\d S}{\d\codeg{B_\A}{p_\A}},\qquad
\ho\codeg{J_{C,\A\B}}{p_\A+p_\B}\equiv\frac{\d S}{\d\codeg{C_{\A\B}}{p_\A+p_\B}}.
\label{JABCdef_nonAbelian}
\end{equation}
The symmetry transformations~\eqref{gauge_nonAbelian} along with the variation of the action~\eqref{dS_nonAbelian} then imply a set of conservation laws, or Ward identities,
\begin{equation}
\begin{split}
\dd\ho\deg{J_{A,i}}{1}&=0,\\
\dd\ho\codeg{J_{B,\A}}{p_\A}&=\k_1\dd\deg{\Om_\A}{p_\A},\\
\dd\ho\codeg{J_{C,\A\B}}{p_\A+p_\B}&=\frac{\k_2}{\k_1}\Bigl[\dd\deg{\Om_\A}{p_\A}\w\ho\codeg{J_{B,\B}}{p_\B}+(-1)^{p_\A p_\B}\dd\deg{\Om_\B}{p_\B}\w\ho\codeg{J_{B,\A}}{p_\A}\Bigr].
\end{split}
\label{Ward_nonAbelian}
\end{equation}
Recall that, in line with footnote~\ref{ftn:diagonal} on page~\pageref{ftn:diagonal}, the right-hand side on the last line has to be augmented with an additional factor of $1/2$ in case that $\A=\B$ and $p_\A=p_\B$ is even.

Finally, as mentioned above, the variation of the action~\eqref{dS_nonAbelian} may be nonvanishing, depending on whether the gauged cohomology generators $\deg{\Om_\A}{p_\A}$ are closed or not. In the latter case, a 't Hooft anomaly is present, which may be understood as arising from a CS theory in the $(D+1)$-dimensional bulk,
\begin{equation}
\deg\I{D+1}=\frac{\k_1}{2\pi\im}\sum_\A(-1)^{D+p_\A}\codeg{B_\A}{p_\A}\w\dd\deg{\Om_\A}{p_\A}.
\label{ID1nonAbelian}
\end{equation}
This in turn leads to a $(D+2)$-form anomaly polynomial that characterizes the anomaly without ambiguities related to its representation in the $D$-dimensional action,
\begin{equation}
\deg\I{D+2}=\frac{\k_1}{2\pi\im}\sum_\A(-1)^{D+p_\A}\dd\codeg{B_\A}{p_\A}\w\dd\deg{\Om_\A}{p_\A}=\frac{\k_1}{2\pi\im}\sum_\A(-1)^{D+p_\A}\codeg{G_\A}{p_\A-1}\w\dd\deg{\Om_\A}{p_\A}.
\label{ID2nonAbelian}
\end{equation}

To see if there actually are any examples of physical systems featuring such a non-Abelian generalization of composite currents, one can follow the line of argument of section~\ref{subsec:examples}. In a theory with two cohomology generators $\deg{\Om_\A}p$ and $\deg{\Om_\B}q$, the exterior product $\deg{\Om_\A}p\w\deg{\Om_\B}q$  will be the Hodge dual of a new conserved current only if $p+q\leq D-1$. Since the degree-one cohomology generators are reserved for Abelian symmetries that were thoroughly discussed in the preceding sections, we need at least one of $p,q$ to be larger than 1 for a qualitatively new example. This leaves us with a single possibility in $D=4$ spacetime dimensions, namely $(p,q)=(1,2)$ up to permutation. This combination of cohomology generators appears, for instance, in the class of theories with a direct product symmetry-breaking pattern, $G\times\gr{U(1)}\to H\times\{\}$, such that $H$ itself has as at least one $\gr{U(1)}$ factor, giving the 2-form cohomology generator~\eqref{cohom2}. The simplest example of symmetry of this type is $\gr{SU(2)}\times\gr{U(1)}\to\gr{U(1)}\times\{\}$, which can be thought of as describing a phase of condensed matter where (anti)ferromagnetism and superconductivity coexist. In this interpretation, the $\gr{SU(2)}$ factor of the symmetry group corresponds to spin, whereas the spontaneously broken $\gr{U(1)}$ factor to electric charge~\cite{Froehlich1993a}.


\section{Summary and discussion}
\label{sec:conclusions}

In this paper, I have outlined a new mechanism whereby a set of mutually independent continuous higher-form symmetries in a given field theory can be deformed into a nontrivial higher-group structure. The basic idea has a very simple formulation in the language of differential forms. A continuous symmetry, 0-form or higher-form, corresponds to a conserved current whose Hodge dual is a closed form. The Hodge duals of all conserved currents of the given theory naturally span a closed algebra under the exterior product. Coupling the whole hierarchy of currents to their respective background gauge fields inevitably leads to a higher-group structure.

Several different incarnations of this mechanism were introduced and analyzed throughout the paper. The one of probably the greatest practical value is the simplified setup of section~\ref{sec:abelian} where only second-order composite currents are taken into account and all the primitive symmetries are assumed to be Abelian. This can however be generalized in different directions with little effort. Thus, section~\ref{sec:abelianhigher} shows how the whole hierarchy of higher-order composite currents can be included, albeit with the simplifying assumption that the given, Noether symmetries are 0-form in addition to being Abelian. Section~\ref{sec:nonabelian} gives up on the latter assumption and shows how one can utilize higher-form topological symmetries arising from the de Rham cohomology of the coset space, defined by the spontaneous breakdown of the given 0-form symmetries.

Here I would like to append a few concluding remarks. First, while the bulk of this paper is formulated using background field methods, the view of higher-form symmetry based on topological operators and extended charged objects is no less interesting. In this regard, the concrete examples worked out in section~\ref{subsec:examples} suggest that quite generally, one can expect the charged object of a composite current to consist of charged objects of the constituent currents, topologically linked to each other. Its charge is in turn defined by integrating (the Hodge dual of) the composite current over a surface that is topologically the Cartesian product of surfaces defining charges of the constituent charged objects.

Second, the whole construction was pitched as being crucially based on the existence of \emph{topological} composite currents, that is currents that are identically conserved and arise from the geometry of the given system rather than from its dynamics. This limitation is, in fact, not essential. In section~\ref{sec:duality}, the dynamical field variables are changed by a duality transformation, which roughly speaking interchanges the role of Noether and topological currents. It is therefore perfectly possible to find the same higher-group structure using composite Noether currents. The realization of the symmetry on the dynamical degrees of freedom then, however, becomes very nontrivial and could hardly be guessed directly, without using duality in the first place.

Third, the notation used throughout the paper suggests that the composite currents are constructed by taking exterior products of (Hodge duals of) \emph{different} primitive currents. This is not really necessary. One can certainly think of topologically conserved composite currents of the type $\ho(\dd\deg\p p\w\dd\deg\p p)$; these are the diagonal, $i=j$ contributions to the action~\eqref{masteraction_Abelian}. The presence of such contributions is however seriously limited by spacetime dimension. Namely, the graded symmetry of the exterior product requires that $p$ be odd, hence such ``diagonal'' composite currents can only exist in $D\geq5$, and the lowest dimension in which they correspond to a higher-form symmetry is $D=6$~\cite{Cordova2020a}.

Finally, let me reiterate that the whole paper is phrased in the language of bosonic low-energy EFT, where all the topological currents, primitive or composite, emerge in the infrared as a consequence of spontaneous breakdown of a given continuous symmetry, responsible for the dynamical NG degrees of freedom. The type of EFT put forward here describes the low-energy dynamics of such systems. One might, however, also be interested in their equilibrium or near-equilibrium thermodynamics. The first attempt to develop a hydrodynamic description of systems with a higher-group symmetry has already been made~\cite{Iqbal2020a}, and the concrete examples presented in this paper might hopefully provide motivation for further efforts in this direction.


\acknowledgments

I am grateful to Yoshimasa Hidaka for introducing me to the world of higher-form symmetries and for useful comments on a draft of the paper. I would also like to thank Sergej Moroz for an illuminating discussion of particle-vortex duality in two-dimensional superfluids. This work has been supported by the grant no.~PR-10614 within the ToppForsk-UiS program of the University of Stavanger and the University Fund.


\bibliographystyle{JHEP}
\bibliography{references}

\end{document}